\documentclass[10pt, conference, compsocconf]{IEEEtran}
\usepackage{times}
\usepackage{subfigure}
\usepackage{graphicx}
\usepackage{amsmath}
\usepackage{bm}
\usepackage{url}
\usepackage{stmaryrd}
\usepackage{balance}
\usepackage{amssymb}
\usepackage{pgfplots}
\usepackage{pifont}
\usepackage[usenames,dvipsnames]{pstricks}
\usepackage{epsfig}
\usepackage{booktabs}
\usepackage{pgffor}
\usepackage{tikz}
\usepackage{multirow}
\usepackage{amsmath}

\makeatletter
\newcommand\xleftrightarrow[2][]{%
  \ext@arrow 9999{\longleftrightarrowfill@}{#1}{#2}}
\newcommand\longleftrightarrowfill@{%
  \arrowfill@\leftarrow\relbar\rightarrow}
\makeatother

\newcommand{\mb}{\mathbf}

\newcommand{\our}{\textsc{All}}
\newcommand{\ouruser}{\textsc{User}}
\newcommand{\ourstation}{\textsc{Station}}
\newcommand{\ourtime}{\textsc{Time}}

\usepackage{algorithm}
\usepackage{algorithmic}

\begin{document}
\title{Bicycle-Sharing System Analysis and Trip Prediction}

\author{

\IEEEauthorblockN{Jiawei~Zhang$^\star$, Xiao Pan$^\dagger$, Moyin Li$^\star$, Philip S. Yu$^\star$$^\ddagger$}
\IEEEauthorblockA{$^\star$University of Illinois at Chicago, Chicago, IL, USA\\
$^\dagger$Shijiazhuang Tiedao University, China\\
$^\ddagger$Institute for Data Science, Tsinghua University, China\\
jzhan9@uic.edu, smallpx@stdu.edu.cn, mli60@uic.edu, psyu@cs.uic.edu}
}

%
\maketitle

\begin{abstract}

Bicycle-sharing systems, which can provide shared bike usage services for the public, have been launched in many big cities. In bicycle-sharing systems, people can borrow and return bikes at any stations in the service region very conveniently. Therefore, bicycle-sharing systems are normally used as a short-distance trip supplement for private vehicles as well as regular public transportation. Meanwhile, for stations located at different places in the service region, the bike usages can be quite skewed and imbalanced. Some stations have too many incoming bikes and get jammed without enough docks for upcoming bikes, while some other stations get empty quickly and lack enough bikes for people to check out. Therefore, inferring the potential destinations and arriving time of each individual trip beforehand can effectively help the service providers schedule manual bike re-dispatch in advance. In this paper, we will study the individual trip prediction problem for bicycle-sharing systems. To address the problem, we study a real-world bicycle-sharing system and analyze individuals' bike usage behaviors first. Based on the analysis results, a new trip destination prediction and trip duration inference model will be introduced. Experiments conducted on a real-world bicycle-sharing system demonstrate the effectiveness of the proposed model.

\end{abstract}


\begin{IEEEkeywords}
Trip Prediction, Bicycle-Sharing System, Mobile Data Mining
\end{IEEEkeywords}
\section{Introduction}\label{sec:introduction}

Bicycle-sharing system refers to a public transportation service system in urban areas offering bicycles for shared use to individuals in a relatively short period of time (about $30-45$ minutes) for free or with very low charges \cite{M11}. In bicycle-sharing systems, people can borrow bikes from stations near them and return the bike to any stations in the city, which can be used as a short-distance trip supplement for private vehicles as well as regular public transportation (e.g., buses and metro trains). Bicycle-sharing system is green and of low carbon, and each bike can be used by several people per day. What's more, due to the widely spread branches and stations available in the city, people can usually borrow and return the bikes very conveniently without wasting time on waiting (needed for the public transportation) or concerns about parking issues in cities (of private vehicles). As a result, bicycle-sharing systems are becoming more and more popular nowadays, which have been adopted in many large cities, e.g., Chicago (Divvy Bike), New York (Citi Bike), San Francisco (Bay Area Bike Share), Washington, D.C. (Capital Bikeshare).

Bicycle-sharing system allows people to borrow bikes with either ``one-day pass'' or ``annual subscribed membership''. ``One-day pass'' is usually preferred by people for temporary usages, e.g., tourist for short-time sightseeing, but the charges per day are slightly higher. Meanwhile, ``subscribed membership'' is a great option for people with frequent travel needs, e.g., office worker and students. Generally, trips completed by one-day pass/membership holders within 30 minutes are included in the pass/membership, but trips longer than $30$-minutes may incur overtime fees. More information about the detailed pricing rules is available at Divvy's official website\footnote{https://www.divvybikes.com/pricing}.



Unlike traditional fixed-route public transportation at pre-scheduled time, services provided by bicycle-sharing systems are more flexible and can meet the daily travel needs of different categories of users. Bicycle-sharing system provides a more microscopic perspective to understand individuals' travel behaviors, which include various aspects about the trips, e.g., trip origin station and start time, as well as trip destination stations and end time. Generally, the travel behaviors of different categories of people with various travel purposes can be quite different. For instance, tourists with one-day pass tend to use the bike to travel among attraction spots, while registered subscribers (like workers and students) mainly travel between companies/schools and homes with the bike. 

Meanwhile, for stations located at different places in the city, the bike usage can be quite skewed and imbalanced \cite{LZZC15}. Some stations that individuals like to borrow bikes from will lack enough bikes for people to check out, while some other stations that people normally return the bikes to will get jammed easily without enough docks for upcoming bikes. To support such a claim, we also analyze the real-world bicycle-sharing system data (to be introduced in Section~\ref{sec:data}), and count the numbers of bikes borrowed from/returned to each stations respectively. According to the analysis results, among all the $474$ stations, $470$ of them have historical usage records: $235$ stations have more bikes being checked out (i.e., \# bikes checked out$>$\# bikes returned), $234$ of them have more returned bikes (i.e., \# bikes checked out$<$\# bikes returned), and only one station (station ID: 449) has balanced usages (i.e., \# bikes checked out$=$\# bikes returned). Therefore, one of the most challenging task for the effective operations of bicycle-sharing systems is to manually shift and rebalance the bikes from the jammed stations to the empty ones. Monitoring the bike usage and inferring the potential destinations of individuals' trips in advance (e.g., at the moment when individuals borrow a bike and start their trips) can help the service providers schedule the manual bike re-dispatch beforehand.


\noindent \textbf{Problem Studied}: In this paper, we propose to predict the potential destination station and arriving time when people start their trips and check out bikes from the origin station at the very beginning. The problem is formally defined as the ``trip prediction'' problem.


\begin{table}[t]
\caption{Properties of the Divvy Dataset}
\label{tab:dataset}
\centering
\begin{tabular}{crr}
\toprule

datasets	&trip	&station\\
\midrule 
2013 Q3-Q4	&759,788	&300	\\
\midrule 
2014 Q1-Q2	&905,699	&300	\\
\midrule 
2014 Q3-Q4	&1,548,935	&300	\\
\midrule 
2015 Q1-Q2	&1,096,239	&474	\\

\bottomrule
\end{tabular}\vspace{-15pt}
\end{table}

The trip prediction problem is an interesting yet important research problem, which is also very challenging to address as individuals' bike trips can be quite complicated and depend on various factors:
\begin{itemize}
\item \textit{Users Composition}: The user composition of bicycle-sharing systems can be quite diverse, which include both (1) long-term registered subscribers and short-term temporary users, (2) male users and female users, as well as (3) young, mid-aged and senior users. The trip prediction problem can be strongly correlated to user categories, and a clear categorization of the bike users can be the prerequisite for addressing the problem.

\item \textit{Temporal Travel Patterns}: Start time of a trip is another important factor that may influence individuals' travel behavior as well as the trip prediction problem. Consider, for example, when a registered member (e.g., a student) borrows a bike in the morning on workdays, it is highly likely that he/she will go to a school for classes. Analyzing the individuals' historical temporal travel patterns will help predict the trips more accurately.

\item \textit{Spatial Travel Patterns}: Besides the time factor, the origin location is another important factor affecting the trip and individuals' travel behaviors. For instance, if a temporary one-day pass holder borrows a bike from a station at the entrance of a sight-seeing trail, he may want to go to the end of the trail. Studying and utilization the historical spatial travel patterns of individuals can help improve the trip prediction performance a lot.
\end{itemize}

To address the trip prediction problem, in this paper, we will analyze the user composition, individuals' temporal and spatial travel behavior patterns of a real-world bicycle-sharing system. Based on the analysis results, we will formulate the trip prediction problem and introduce new models to infer both the trip destination station and trip duration.

The remaining part of the paper is organized as follow. In Section~\ref{sec:data}, we first introduce the Divvy bicycle-sharing system dataset and give some basic statistical information about the dataset. The user decomposition of the bicycle-sharing system Divvy is available in Section~\ref{sec:user}. Individuals' temporal and spatial travel patterns are studied in Section~\ref{sec:temporal} and Section~\ref{sec:spatial} respectively. Based on the analysis results, we formulate the trip prediction problem in Section~\ref{sec:formulation} and introduce the trip prediction model in Section~\ref{sec:method}, which is evaluated in Section~\ref{sec:exp}. Finally, we discuss the related works in Section~\ref{sec:related} and conclude the paper in Section~\ref{sec:conclusion}.

\section{Divvy Dataset Description} \label{sec:data}

Before analyzing the individuals' travel behaviors, we will introduce the dataset about a real-world bicycle-sharing system first in this section. The dataset used in this paper is about the Divvy bicycle-sharing system initially launched in the Chicago city on June 28, 2013. At the very beginning, Divvy had about $750$ bikes at $75$ stations (operating in an area spanning from the Loop north to Berwyn Avenue, south to $59th$ Street, west to Kedzie Avenue, and east to the Lake Michigan coast). A quick expansion has been made at early 2015, and Divvy now operates $4,760$ bicycles at $474$ stations (in an area bounded by $75th$ Street on the south, Touhy Avenue on the north, Lake Michigan on the east, and Pulaski Road on the west).

The Divvy bicycle-sharing system datasets are public and new datasets are released every two quarters, which can be downloaded at its official website\footnote{https://www.divvybikes.com/data}. We downloaded the Divvy bicycle-sharing system data on November 2, 2015, which contains $4$ separate datasets time ranging from the middle of 2013 to the middle of 2015 respectively. The downloaded datasets include the complete historical trip records as well as the station information, whose statistical information and detailed descriptions are available in Table~\ref{tab:dataset} and as follows.

\begin{figure}[t]
\centering
    \begin{minipage}[l]{0.6\columnwidth}
      \centering
      \includegraphics[width=\textwidth]{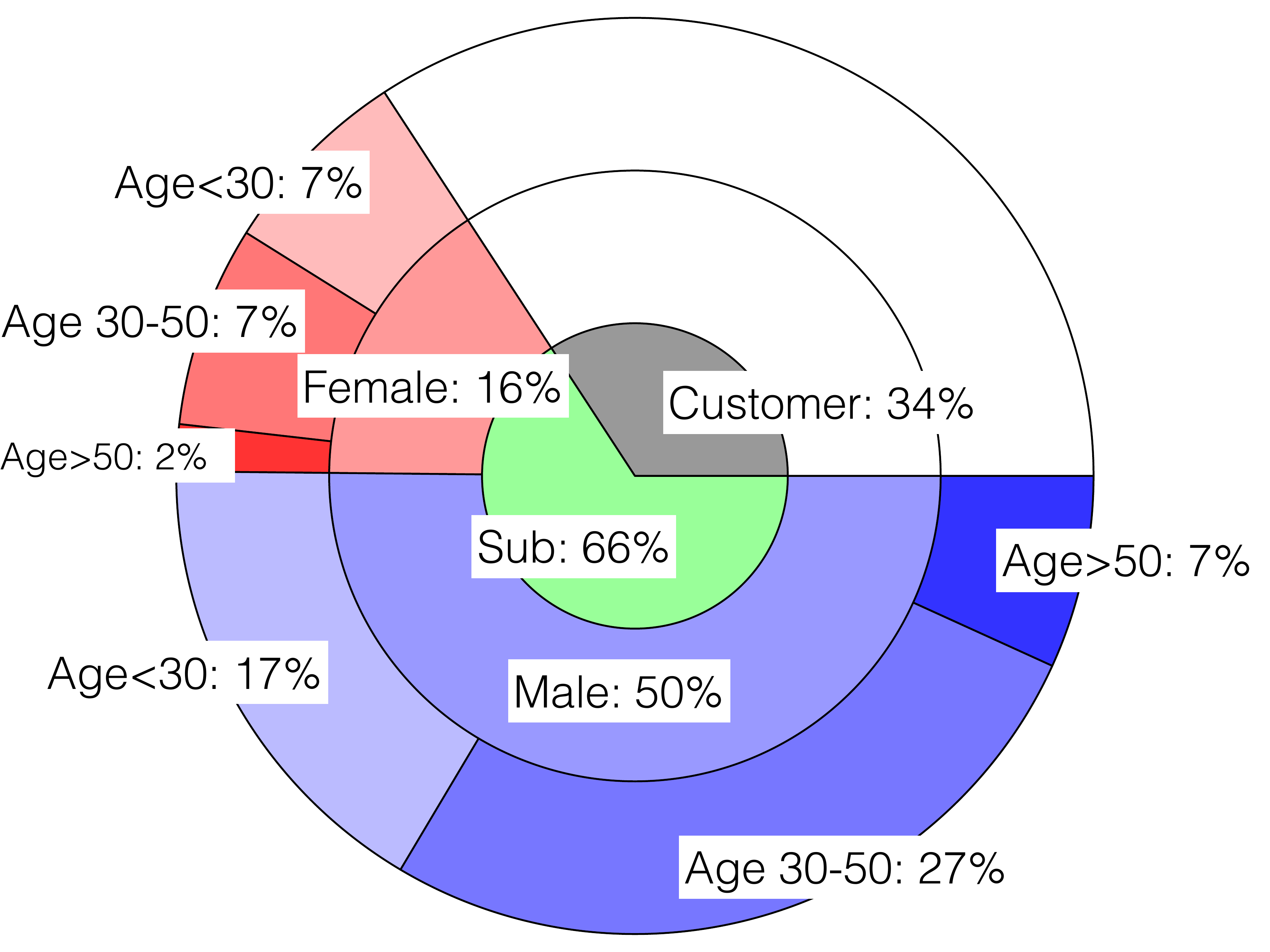}
    \end{minipage}
  \caption{Percentages of trip belong to different categories of users (Red: Female Subscribers; Blue: Male Subscribers; Green: Subscribers; Gray: Customers).}\label{fig:user}\vspace{-15pt}
\end{figure}

\begin{figure*}[t]
\centering
    \begin{minipage}[l]{1.8\columnwidth}
      \centering
      \includegraphics[width=\textwidth]{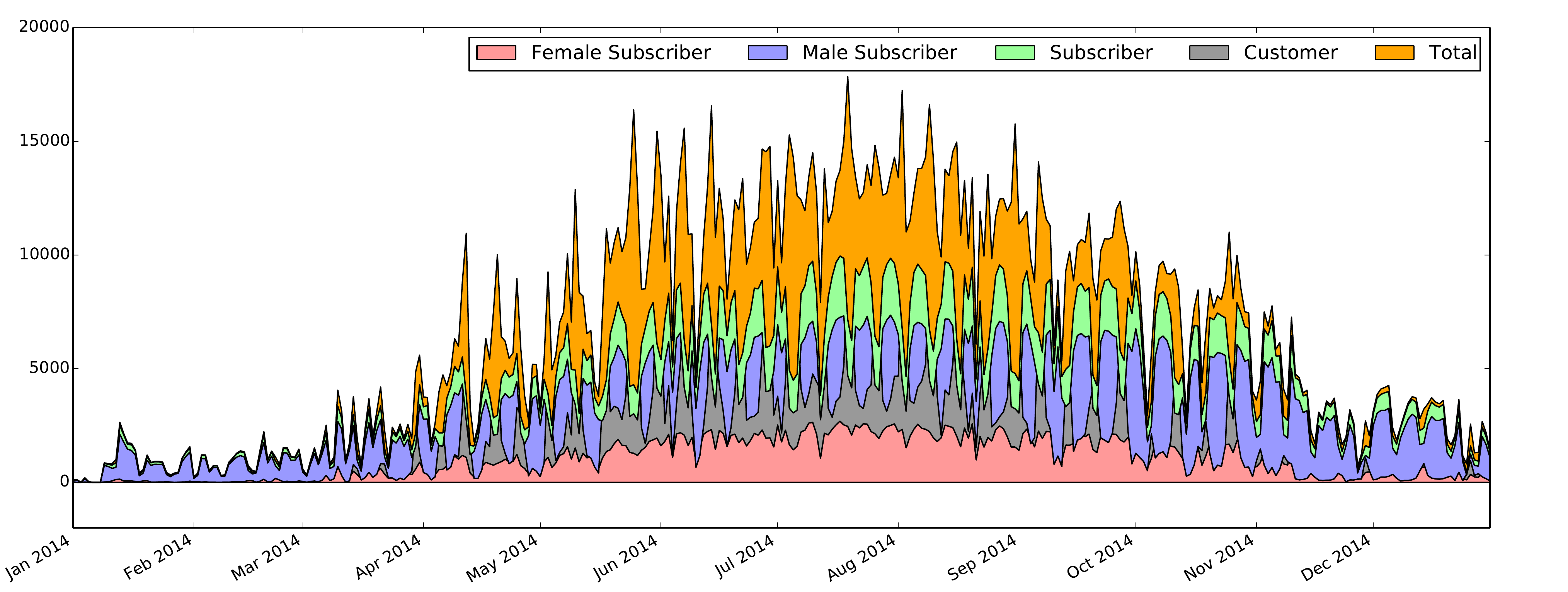}
    \end{minipage}
  \caption{Statistics of trips on each day of the 2014 year (X axis: each day of 2014; Y axis: number of trips in one day).}\label{fig:2014_stat}\vspace{-15pt}
\end{figure*}

\begin{itemize}

\item \textit{Trip}: Each trip record in the datasets has a unique ID. From the trip record data, we can know the trip start and end time as well as the corresponding origin and destination bike stations. The trip record also indicates whether the user is an annual membership holder or just an one-day pass holder, who are called the ``subscriber'' and ``customer'' respectively. For the annual membership subscribers, the trip record data also includes their gender and birth year information, which is helpful for categorizing the users (into male vs female, as well as youth vs senior) and allows us to study the bike-usage behaviors of different categories of people.

\item \textit{Station}: For each station, we can know its ID, name as well as its specific location, which is represented as a (latitude, longitude) coordinate pair in the dataset. At stations, bikes are locked at the docks and the numbers of docks available at the stations are called the station capacities, which are also available in the datasets.

\end{itemize}

As shown in Table~\ref{tab:dataset}, the numbers of trips in these $4$ separate datasets are 2013 Q3-Q4: $759,788$; 2014 Q1-Q2: $905,699$; 2014 Q3-Q4: $1,548,935$; 2015 Q1-Q2: $1,096,239$ respectively. Generally, the Chicago people like to use the Divvy bike a lot and, on average, $179,610$ trips were taken in each month during the past two years. Meanwhile, the number of stations doesn't change in the first $3$ datasets (which are all $300$), and increases to $474$ in the last dataset because of the scheduled expansions at the beginning of 2015.

In the following sections, we will study the datasets in great detail to analyze the user composition, individuals' temporal travel patterns, and spatial travel patterns respectively. Based on the analysis results, we will introduce the trip prediction problem and the model to address the problem.



\section{Divvy User Composition}\label{sec:user}

To predict the trips taken by users in bicycle-sharing systems, we need to understand the composition of people using the bike at first. By studying the historical trip record data, we count up the numbers of trips taken by ``customers'' and ``subscribers'' respectively among the Divvy users and the statistical results are shown in Figure~\ref{fig:user}.


From Figure~\ref{fig:user}, we observe that the majority of trips are actually taken by the ``subscribers'' (i.e., the green area marked with ``Sub''), which account for about $66\%$ in the total trips, while those finished by the ``customers'' (i.e., the gray area) account for $34\%$ in all. 



No extra information is available for the ``customers'', as they just buy one-day pass and no personal information is recorded. Meanwhile, for the ``subscribers'' with formal membership registrations, we can know more (e.g., gender and age) about them and can further study their compositions.


As shown in Figure~\ref{fig:user}, the ``subscribers'' area is further divided into the ``male'' and ``female'' subscribers. Among all these Divvy bike trips, ``male subscribers'' (i.e., the blue area) finish about $50\%$ of them, and ``female subscribers'' (i.e., the red area) have taken $16\%$ of the trips. 





In addition, we also count the trips finished by people belonging to $3$ different age groups, which include young people: age$<$30; mid-aged people: 30$\le$age$<$50; and senior people: age$\ge$50, which are denoted by the red/blue color of different saturations in Figure~\ref{fig:user}. From the result, we observe that among the $50\%$ bike trips finished by the ``male subscribers'', the ratio of trips taken by the young, mid-aged and senior people account for $17\%$, $27\%$ and $7\%$ respectively. Meanwhile, the trip finished by the female subscribers belonging to these $3$ groups are $7\%$, $7\%$, and $2\%$ respectively. Therefore, the Divvy bike is preferred and frequently used by the young and mid-aged people, who together finish about $58\%$ of the total trip.

In summary, based on the analysis results, we can partition the users into several categories (e.g., ``customers'' vs ``subscribers'', ``male'' vs ``female'', young vs mid-aged vs senior). In the following sections, we will study the temporal and spatial travel patterns of different categories of users in detail.


\begin{figure*}[t]
\centering
\subfigure[Each month in a year]{ \label{fig:month}
    \begin{minipage}[l]{.55\columnwidth}
      \centering
      \includegraphics[width=\textwidth]{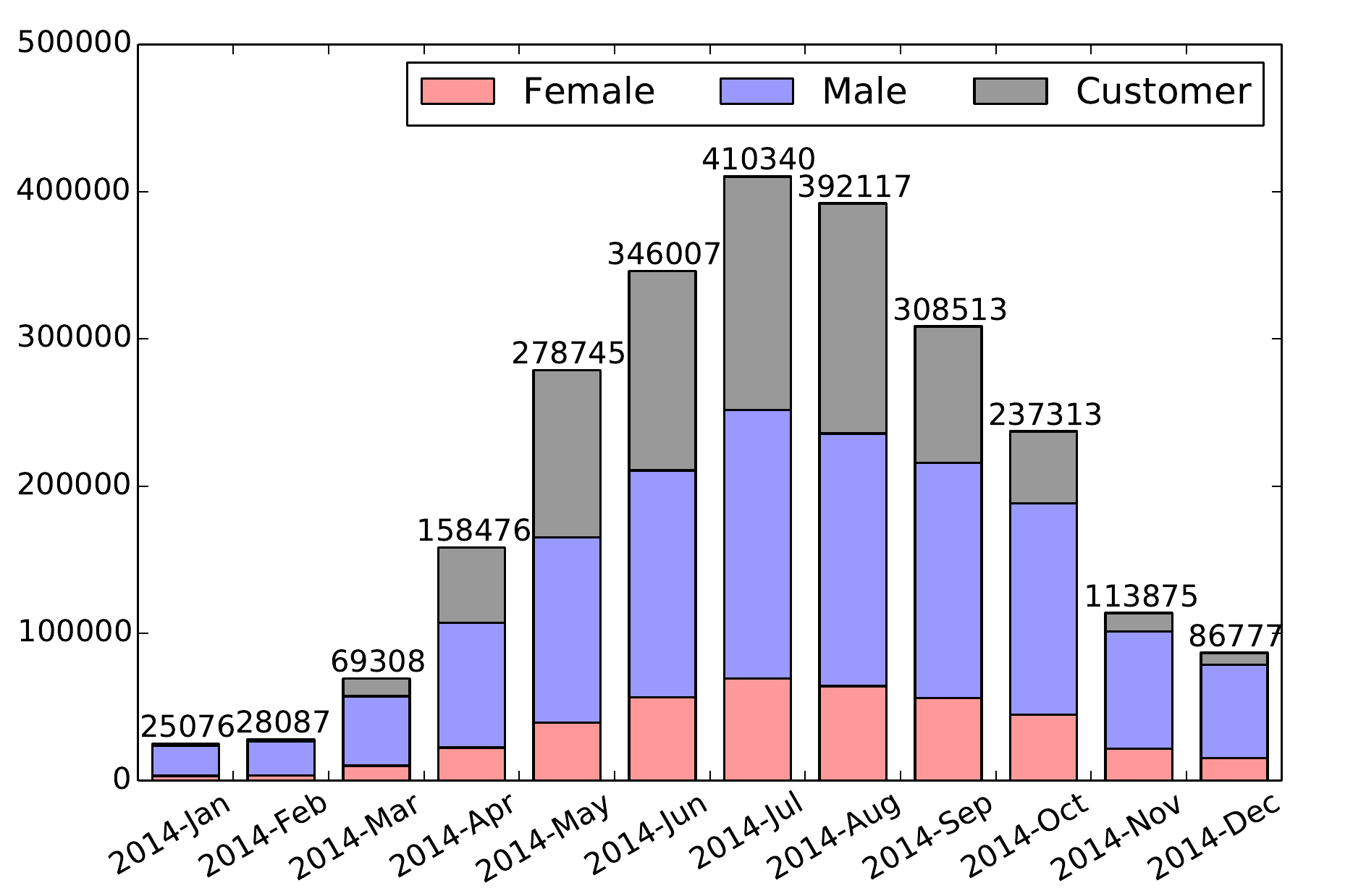}
    \end{minipage}
  }
\subfigure[Each weekday in a week]{ \label{fig:weekday}
    \begin{minipage}[l]{.6\columnwidth}
      \centering
      \includegraphics[width=\textwidth]{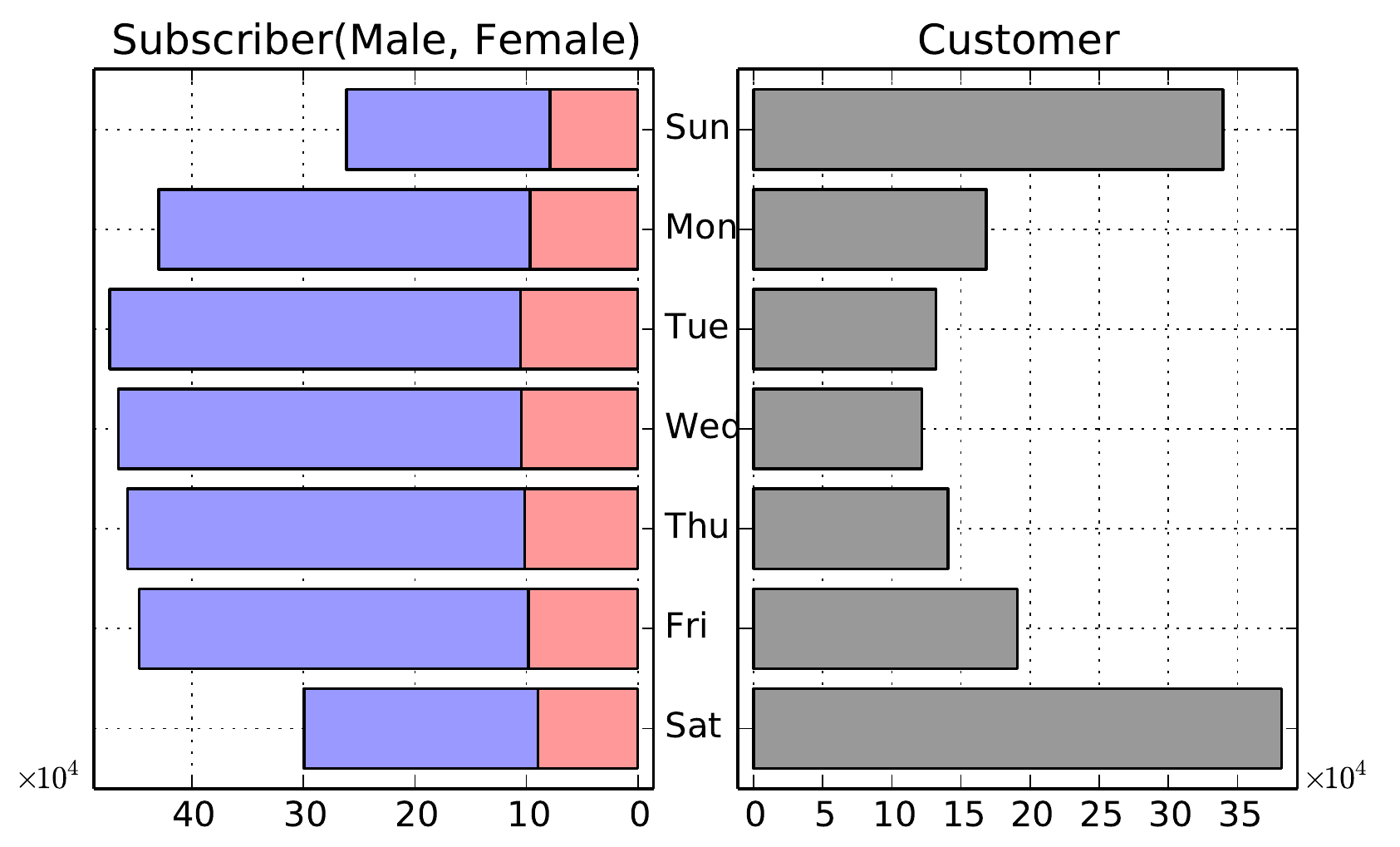}
    \end{minipage}
  }
\subfigure[Each hour in a day]{ \label{fig:hour}
    \begin{minipage}[l]{.55\columnwidth}
      \centering
      \includegraphics[width=\textwidth]{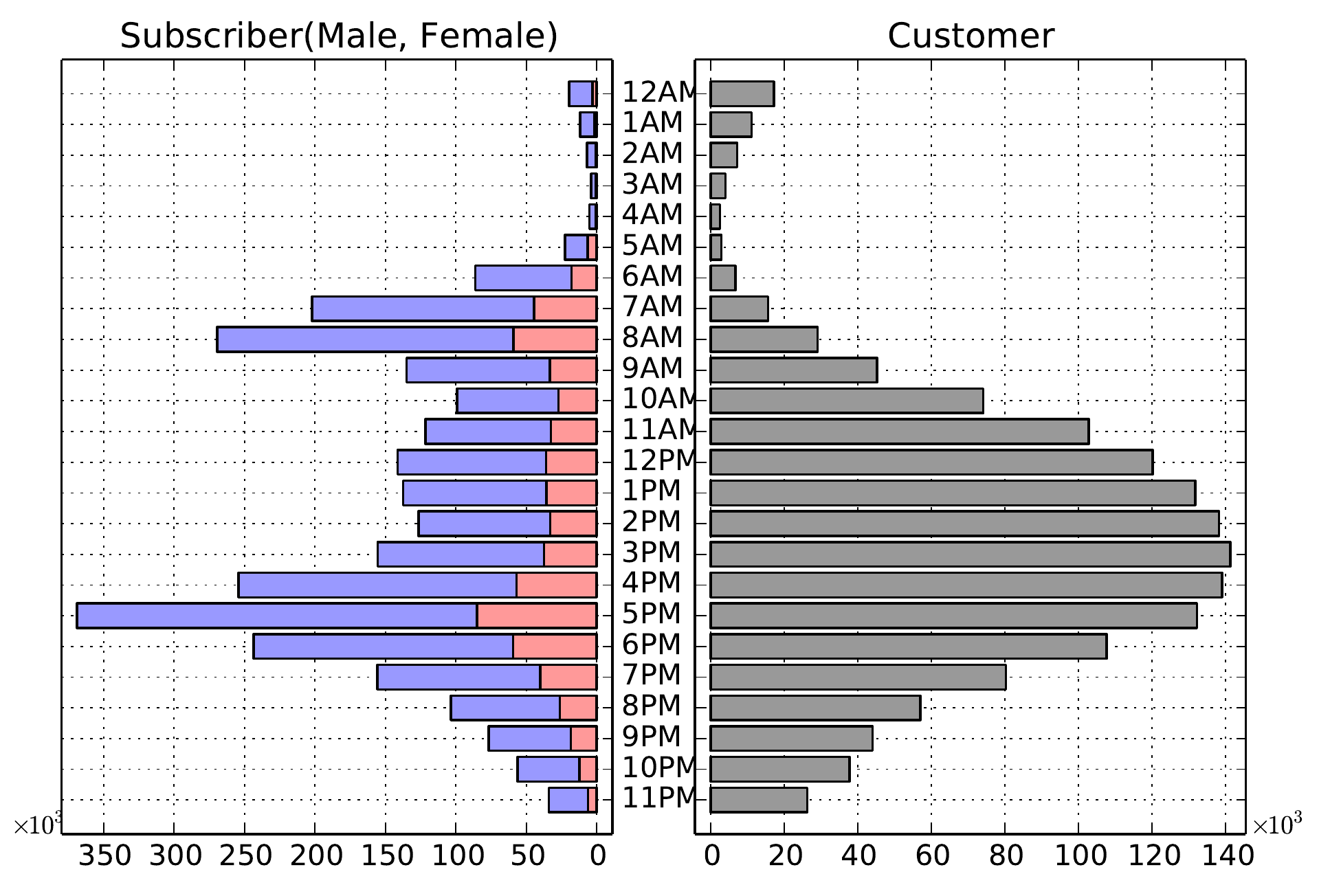}
    \end{minipage}
  }\vspace{-10pt}
  \caption{Cyclical patterns in a year, a week, and a day.}\label{fig:weather}\vspace{-15pt}
\end{figure*}

\section{Temporal Travel Patterns}\label{sec:temporal}

In this section, we will study individuals' temporal travel patterns with the Divvy bike. We will first study and analyze the Divvy trip taken on each day in the 2014 year, from which individuals' cyclic travel patterns can be observed. For different cycle lengths (one year, one week and one day), we analyze the distribution of individuals' bike usages in each cycle. Finally, we also study the time durations of historical bike trips in the dataset.

\subsection{Trip Temporal Distribution Overview}


The Divvy bicycle-sharing system provides bike sharing services throughout the whole year. To have a look at the bike usages within a year in the dataset, we count up the trip records on each day in 2014 taken by ``customers'', ``male subscribers'' and ``female subscribers'' respectively, whose results are available in Figure~\ref{fig:2014_stat}.


From Figure~\ref{fig:2014_stat}, we observe that people use the Divvy bike everyday, but the majority of the trips concentrate within the months ranging from April to November, and the number of trips taken during the winter seasons is quite limited (we will further study the monthly bike usages in 2014 in Section~\ref{subsec:temporal}). Such a phenomenon can be correlated to the weather in Chicago and, to support such a statement, we also check the historical weather within the Chicago area in 2014 from Weather Underground\footnote{http://www.wunderground.com}. According to the historical weather data in Chicago, the average temperatures during January and February of 2014 were below $20^{\circ}F$, and the average temperature in November and December of 2014 were below $40^{\circ}F$ respectively. Meanwhile, over 20 days snowed in January 2014, and the numbers of snowing days during the February, November and December were all larger than 10. In this kind extreme weather conditions, travelling by bike is almost infeasible. Meanwhile, as the weather gets better, Divvy bike usage increases steadily.




Besides the weather reasons, some other factors can also influence the Divvy bike usages, like various events celebrated in Chicago. For instance, from Figure~\ref{fig:2014_stat}, we observe that people's bike riding activities reach the peak on July 19-20, 2014 (Saturday and Sunday) in Chicago. According to Chicago event schedule\footnote{http://www.choosechicago.com/articles/view/CHICAGO-EVENTS-FESTIVALS-2014-CALENDAR-HIGHLIGHTS/1243/}, at the same time, various events were taken place at Chicago including the ``Pitchfork Music Festival'' (tens of thousands of music fans are involved and gathered together), ``Taste of River North'', ``Chicago Craft Beer festival'', etc. To attend these celebration festivals, Divvy bikes with no worries about the parking issues are the ideal travel options for people. Viewed in this perspective, the Divvy bike riding activities are also correlated with the offline events.

In addition, by investigating the bike sharing activity changing trends in Figure~\ref{fig:2014_stat}, we observe that their activity patterns follow cyclical fluctuations. We will give the analysis about people's cyclical Divvy bike usages in the following section.

\subsection{Cyclical Travel Patterns}\label{subsec:temporal}



In this part, we will analyze the cyclical fluctuations of people's bike usages in one year, one week and one day respectively.

\subsubsection{Cyclical Pattern in One Year}


In Figure~\ref{fig:month}, we count the trip numbers of each month in the 2014 year taken by ``customers'', ``male subscribers'', and ``female subscribers'' respectively, and the total number is also marked on top of the bars. Similar to the observations discussed for Figure~\ref{fig:2014_stat}, most of the trips are done during April until October in the year 2014, due to the weather reasons, like temperature and snow precipitations. 

In addition, we observe that the trips taken by the ``customers'' mostly appear during April until October, and they rarely use the Divvy bike during the winter and early spring seasons. To support such a claim, we calculate the ratio of trips taken by ``customers'' against the total number of trips in each month of 2014. The ratio during January and February of 2014 is below $5\%$, and around $10\%$ in November and December. However, the ratio increases rapidly to over $40\%$ during May to August (i.e., the summer season). Therefore, the ``customers'' should be different from the regular long-term ``subscribers'', and they can be temporary visitors to Chicago.



\subsubsection{Cyclical Pattern in Each Week}


From Figure~\ref{fig:2014_stat}, we observe the cyclical fluctuations in the bike usages and the cycle length takes about $\frac{1}{4}$ month, i.e., one week. In this part, we will check whether there is cyclical pattern in each week or not for both ``customers'' and ``subscribers''.

In Figure~\ref{fig:weekday}, based on the whole dataset, we show the number of trips finished by ``customers'', ``male'' and ``female'' subscribers respectively on each weekday from Sunday to Saturday. Based on the results, we observe that ``customer's'' bike usage pattern is totally different from that of ``subscribers''. ``Subscribers'' mainly use the Divvy bike during the weekdays from Monday to Friday, and their usages on weekends (i.e., Sunday and Saturday) drop a lot. For instance, the number of trips taken by the ``subscribers'' on Tuesday is $473,957$ in all, but the number drops to $261,321$ on Sunday, which decreases almost by $45\%$. However, the Divvy bike usages for the ``customers'' follow a totally different pattern: ``customers'' tend to use the Divvy bike more often on weekends than weekdays. For instance, the number of trips taken on weekends (Saturday and Sunday) by ``customers'' is $721,073$, which account for $49\%$ of their total trips.


\subsubsection{Cyclical Pattern in Each Day}


Besides the annual and weekly patterns, we also wonder whether people's bike usages have daily cyclical patterns or not. In Figure~\ref{fig:hour}, we divide each day into 24 hours and count the trips taken at each hour, where the trips time denotes their starting time. 


From Figure~\ref{fig:hour}, we can observe that most of the Divvy trips are taken during the daytime from 6AM to 7PM for both ``customers'' and ``subscribers''. However, the activity patterns for ``customers'' and ``subscribers'' are totally different: (1) ``subscribers'' mostly ride the Divvy bike during 7AM-9AM and 4PM-6PM; but (2) ``customers'' use the Divvy bike during the daytime from 10AM-6PM. The peak Divvy usage hours of ``subscribers'' happen to be most employees' and students' commute rush hours, and the ``subscribers'' mainly use Divvy for (at least part of) their workplace-home commutes. Meanwhile, the peak Divvy usage hours of ``customers'' concentrate round the daytime, especially the visitors' sightseeing hours.

\subsection{Trip Time Duration}\label{subsec:length}

\begin{figure}[t]
\centering
    \begin{minipage}[l]{0.7\columnwidth}
      \centering
      \includegraphics[width=\textwidth]{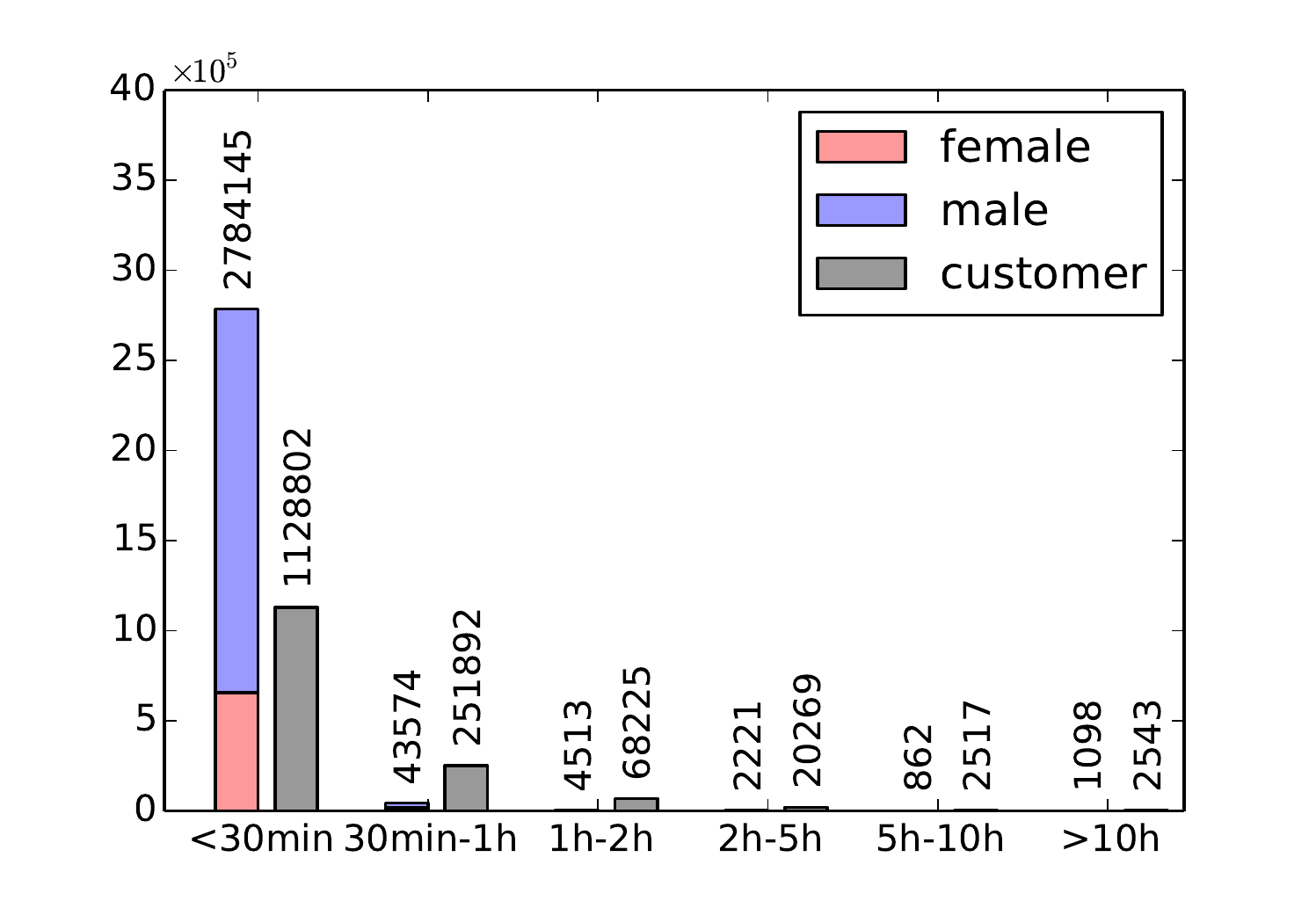}
    \end{minipage}
  \caption{Trip Duration (X axis: trip time duration, Y axis: \# trips).}\label{fig:trip_length}\vspace{-15pt}
\end{figure}


In addition to the cyclical bike usage patterns, we are also interested in the time durations of Divvy bike trips. To study the bike usages in the real-world, we calculate the average trip time duration in the whole dataset to be $17.76$ minutes. To study the detailed distributions of trip time length, in Figure~\ref{fig:trip_length}, we partition the trip length into $6$ bins: \{$<$30 minutes, 30 minute-1 hour, 1 hour-2 hours, 2 hours-5 hours, 5 hours-10 hours, $>$10 hours\} and count the number of trips belonging each time bin. From Figure~\ref{fig:trip_length}, we can observe that the number of trips which are shorter than $30$ minutes ridden by ``subscribers'' and ``customers'' are $2,784,145$ and $1,128,802$, which together accounts for $90.77\%$ of the total bike trips. In other words, the majority of users will return the bike within the free-ride time (i.e., $30$ minutes) and don't want to pay the over-time charges. However, about $397,714$ bike trips are still longer than $30$ minutes, the majority of which are taken by ``customers''. The number of over-time trips ridden by ``customers'' is $345,446$, and accounts for $86.85\%$ of the total over-time trips. 





\section{Spatial Travel Patterns}\label{sec:spatial}

\begin{figure}[t]
\centering
    \begin{minipage}[l]{0.65\columnwidth}
      \centering
      \includegraphics[width=\textwidth]{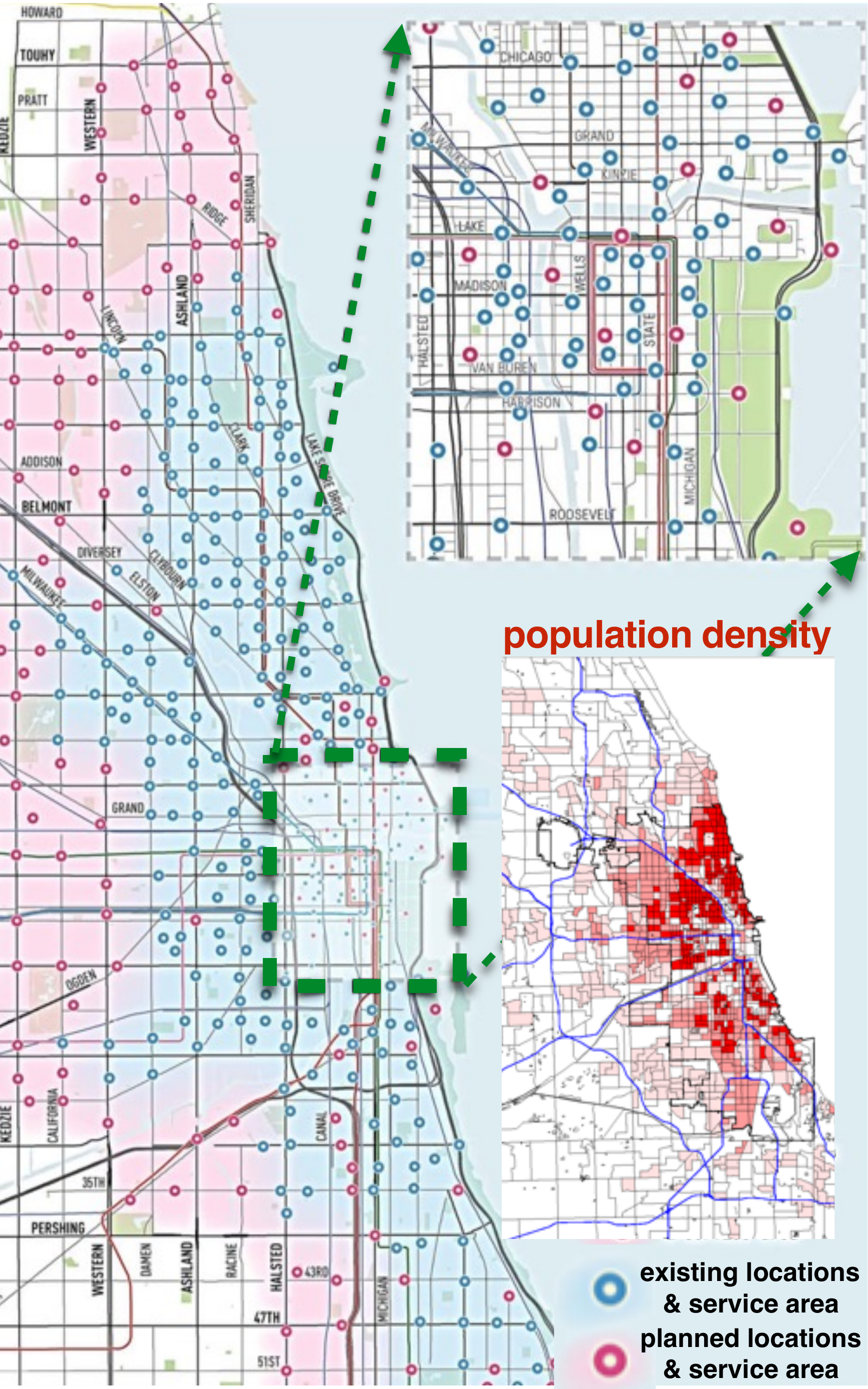}
    \end{minipage}
  \caption{Divvy station distribution and population distribution at Chicago \cite{chicago}.}\label{fig:station}\vspace{-15pt}
\end{figure}

In this section, we will study the individuals' spatial travel patterns. We will first introduce the overall distributions of Divvy bike stations in the Chicago city. Next, we will analyze the geo-distances of trips taken by different categories of people, and study the top 10 stations that individuals frequently start and end their trips at, as well as the station pairs that people usually travel between.

\subsection{Station Spatial Distribution Overview}

In Figure~\ref{fig:station}, we show the distribution of the Divvy stations at the Chicago city, where the blue area and blue dots are the existing service region and the existing Divvy station locations. Due to the vast travel demands from the public, Divvy is expanding its service region to broader areas by adding new stations to both new and existing service regions, i.e., the red area and the red dots. By comparing the number of stations in the existing and planned service regions (i.e., the blue and red areas), we observe the station distribution is denser in the blue region, which also corresponds to the densely populated area at the Chicago city. The most prosperous area at Chicago should be the Loop area, which is also the region that the Divvy bike was initially launched at. To have a clear view about the stations available at the Loop area, in Figure~\ref{fig:station}, we also zoom in the area (i.e., marked in the green dashed square), from which we can observe divvy stations within the Loop region is extremely dense and many new stations are to be added. 



\subsection{Trip Geo-graphical Distance}\label{subsec:distance}

\begin{figure}[t]
\centering
    \begin{minipage}[l]{0.65\columnwidth}
      \centering
      \includegraphics[width=\textwidth]{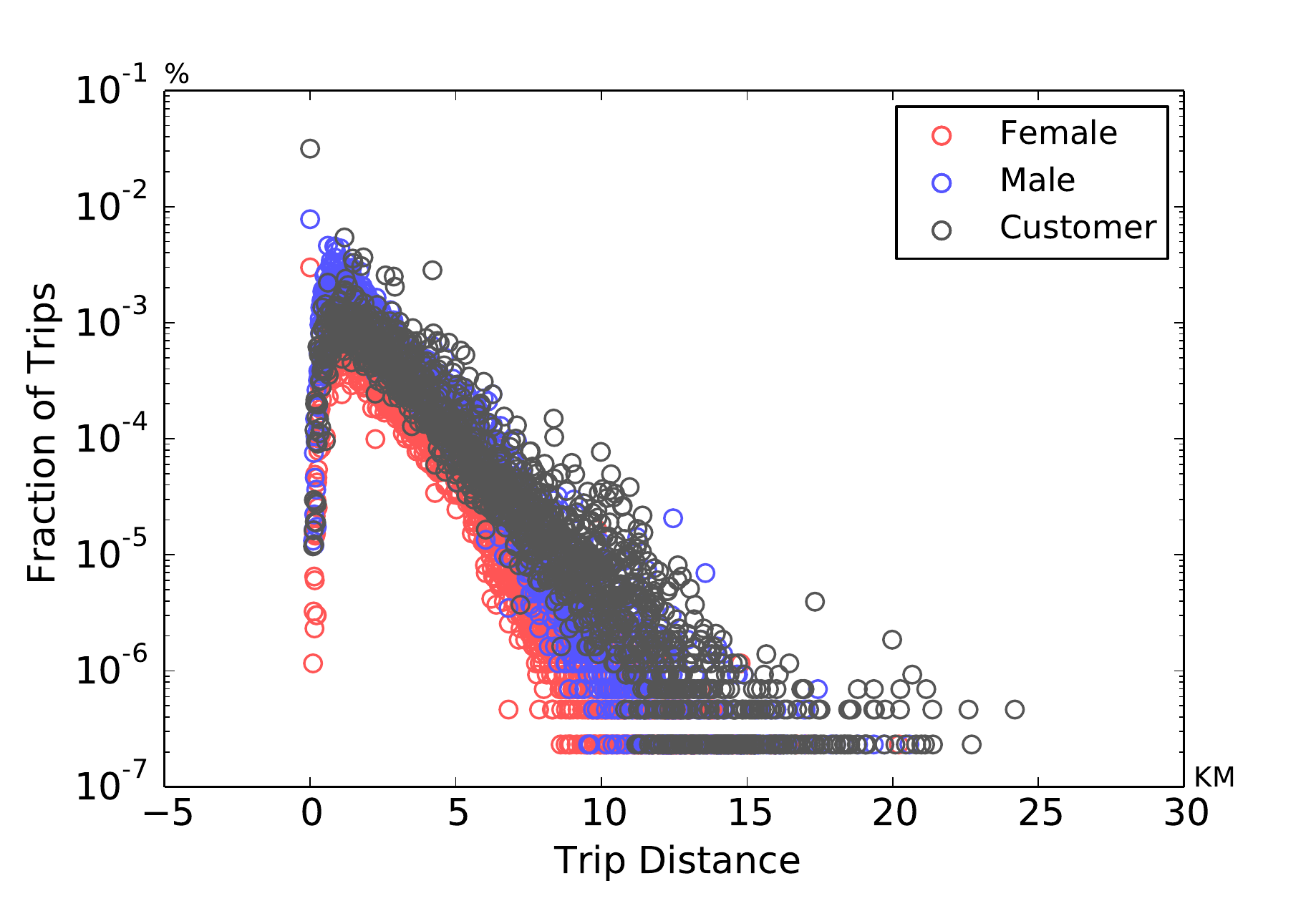}
    \end{minipage}
  \caption{Trip Geo-Distance (X axis: geo-distance (unit: KM), Y axis: fraction of trips).}\label{fig:trip_distance}\vspace{-18pt}
\end{figure}

For bike trips taken by different categories of people among the stations, their distances can vary a lot. In Figure~\ref{fig:trip_distance}, we show the distributions about the distance (in kilometers (KM)) of trips taken by ``customers'', ``male subscribers'' and ``female subscribers'' respectively. Here, \textit{Manhattan Distance} \cite{WM97} is used as the distance measure, as the roads in Chicago are very similar to those in Manhattan. We observe that the distribution curves don't follow the power law distribution \cite{CSN09} exactly, where the majority of trips are within distance about $0.5-5$KM, while those shorter than $0.5$KM or longer than $5$KM only account for a very small proportion. What's more, based on the historical trip data, we calculate the average distance of trips ridden by ``customers'' and ``subscribers'' (both male and female subscribed users) to be $2.12$KM and $1.91$KM respectively. In other words, trips finished by the customers are slightly longer.

\subsection{Top 10 Trip Origin Stations}

\begin{figure}[t]
\centering
    \begin{minipage}[l]{0.8\columnwidth}
      \centering
      \includegraphics[width=\textwidth]{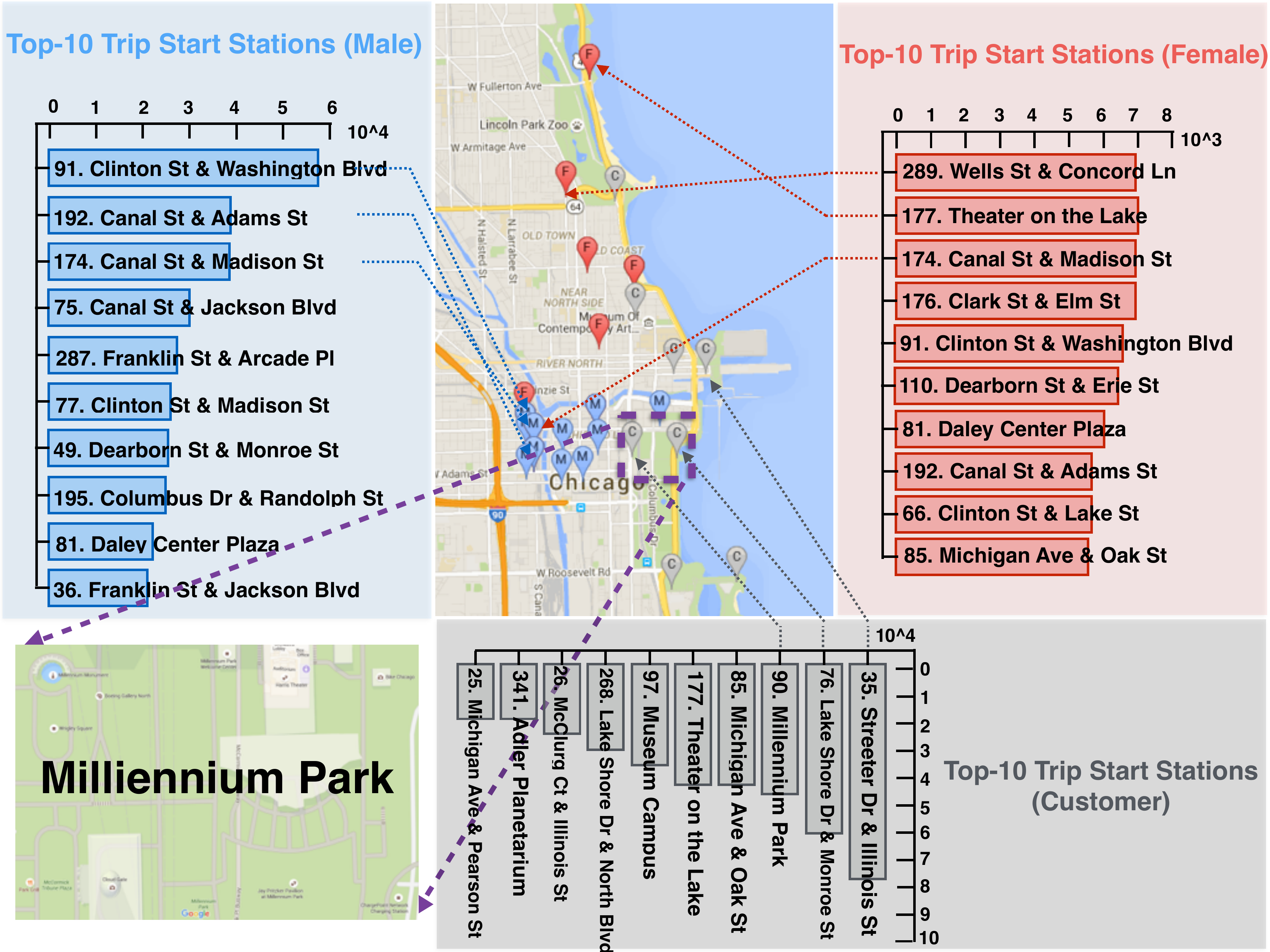}
    \end{minipage}
  \caption{Top 10 trip origin stations.}\label{fig:start}\vspace{-15pt}
\end{figure}

The trip origin stations that people frequently start their trips from (i.e., the trip origin stations) can be quite different. In Figure~\ref{fig:start}, we show the top 10 Divvy bike stations that ``customers'', ``male subscribers'' and ``female subscribers'' usually start their trip from, which include the ranked station list with station IDs and station names. In addition, we also give the number of trips starting from the stations (i.e., the length of the bars), as well as their specific coordinates on the map. Different categories of users can also have certain overlaps in the trip start stations, and some of the overlapped station markers are covered and not shown. For instance, Station 174 serves as the top $3rd$ frequently visited station for both male and female subscribers, but the icon corresponding to the ``female subscribers'' is hidden by that of the ``male subscribers'' on the map.


From the results, we observe that ``male subscribers'' mainly use the Divvy bikes from the loop area; ``female subscribers'' normally use the bike from the northern residential area; and ``customers'' mostly use the bike for sightseeing at attraction spots along the Lake Michigan coast. For instance, (1) the top ranked site that ``male subscribers'' frequently borrow the bike to start their trips is ``Station 91'', and it is located close to the loop area and is next to many sites of very large population flow, which include several important transportation hubs in Chicago (i.e., Chicago Union Station and Ogilvie Transportation Center), apartment buildings, as well as office buildings. (2) The top ranked site that ``female subscribers'' usually start their trips from is ``Station 289'', and it is located at the northern part of the Chicago city, which is a rich and safe residential area in Chicago. And (3) the top ranked start station for ``customers'' is ``Station 35'', which is located near the famous Chicago landmark ``Navy Pier'', and their $2nd$ and $3rd$ top stations are located next to the ``Millennium Park'', which is also a famous tourist site in Chicago.

\subsection{Top 10 Trip Destination Stations}

\begin{figure}[t]
\centering
    \begin{minipage}[l]{0.8\columnwidth}
      \centering
      \includegraphics[width=\textwidth]{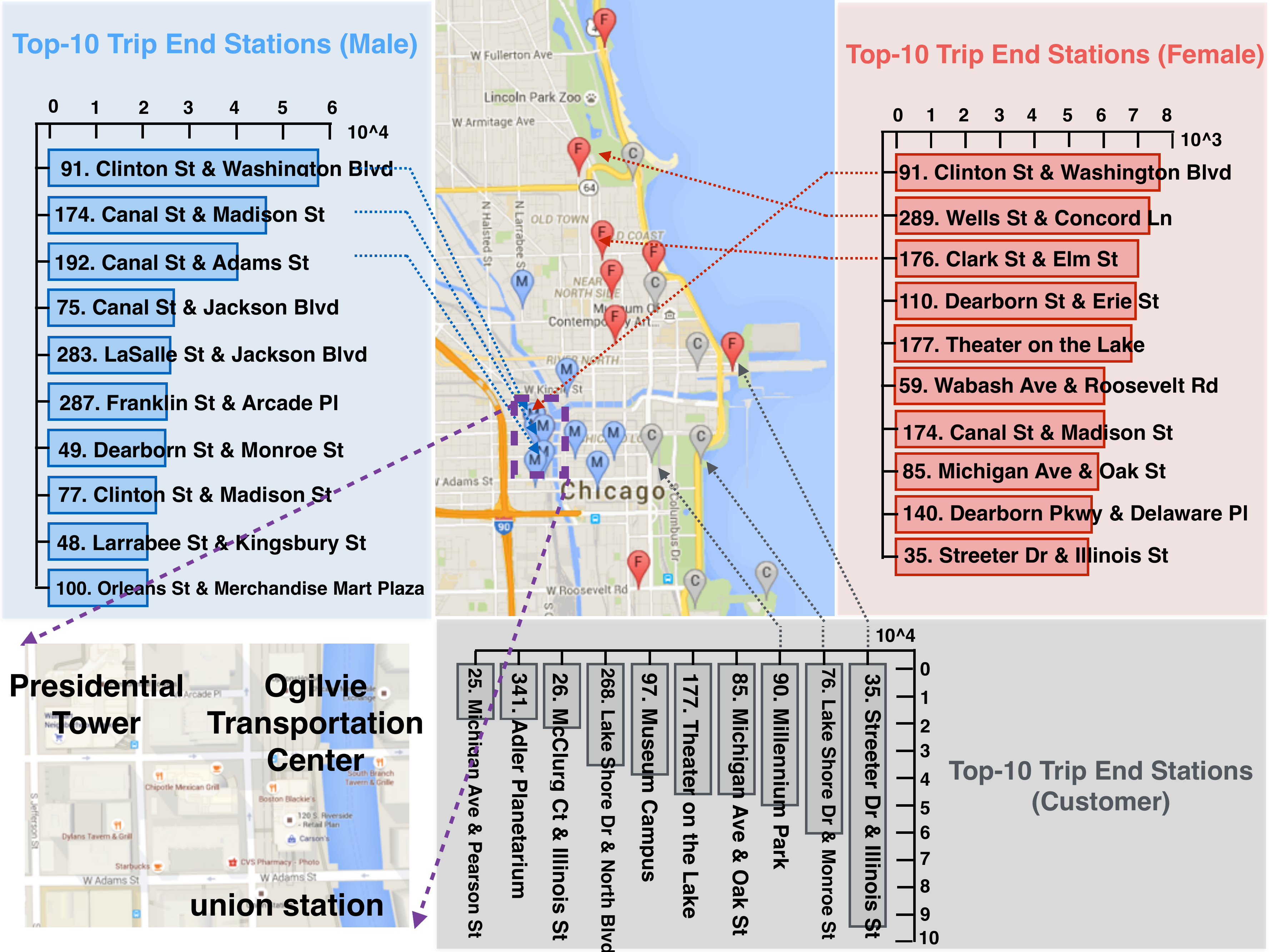}
    \end{minipage}
  \caption{Top 10 trip destination stations.}\label{fig:end}\vspace{-15pt}
\end{figure}

Besides the start stations, we also show the top 10 stations that people usually end their trips, i.e., the destinations of their trips, and the results are shown in Figure~\ref{fig:end}. 

From Figure~\ref{fig:end}, we observe that the top 10 frequently visited stations of ``male subscribers'', ``female subscribers'' and ``customers'' are quite similar to those shown in Figure~\ref{fig:start}. For instance, 7 of the top ranked end stations of ``male subscribers'' and ``female subscribers'' have ever appeared in the corresponding origin station ranking list, and the top 10 end stations of ``customers'' are even identical to their origin stations as shown in Figure~\ref{fig:start}, but the trip numbers (i.e., the bar width) are different. The potential explanation for such a phenomenon can be that the bike trips for many users are normally bi-directional. In other words, if they ride from station A to station B for certain events (which can be ``go to workplaces'' for subscribed users, or ``sightseeing'' for customers), then they may also need to ride back after the events. Therefore, station A and station B both serve as the origin and destination stations respectively in these two trips. 


\subsection{Top 5 Frequently Traveled Station Pairs}

\begin{figure}[t]
\centering
    \begin{minipage}[l]{0.8\columnwidth}
      \centering
      \includegraphics[width=\textwidth]{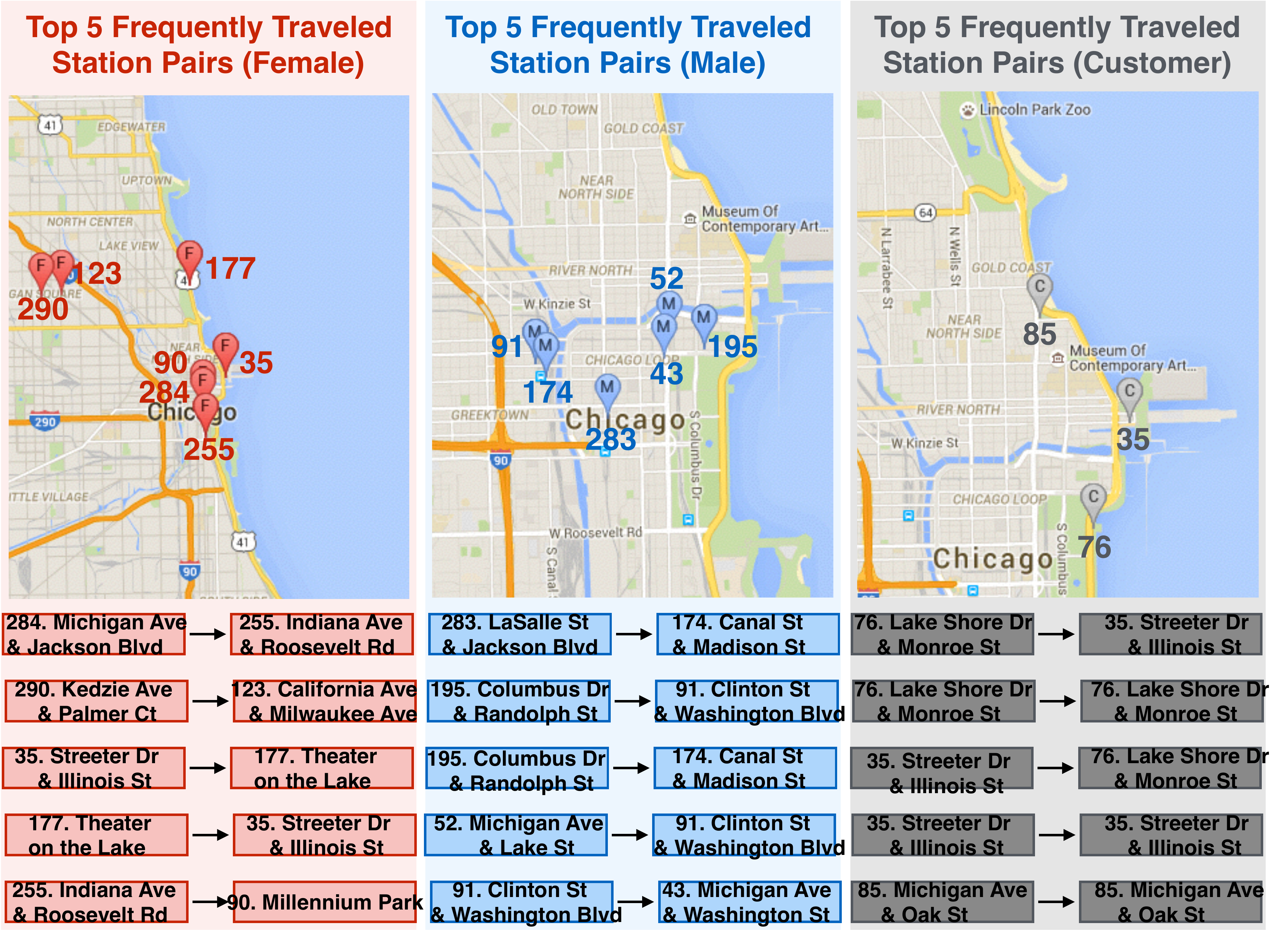}
    \end{minipage}
  \caption{Top 5 most frequently traveled stations.}\label{fig:frequent}\vspace{-15pt}
\end{figure}

Generally, different Divvy bike trips are for different purposes, and the purpose can be captured more clearly by considering the origin and destination stations at the same time. For example, if the bike trip departs from residential region and the destination is a campus, the rider is likely to be a student and uses Divvy bike to commute from home to schools; while if the trip origin and destination stations are both attraction sites, then the rider mainly uses the Divvy bike for sightseeing. Motivated by this, we show the top 5 frequently traveled station pairs of ``male subscribers'', ``female subscribers'' and ``customers'' in Figure~\ref{fig:frequent}, where the origin and destination stations are listed and marked on the map.

From Figure~\ref{fig:frequent}, we observe that the top ranked Divvy station pair for ``male subscribers'' is ``Station 283 $\to$ Station 174'', where station 283 is at the Chicago loop region (i.e., the Chicago city center area full of office buildings) and station 174 is just next to the ``Ogilvie Transportation Center''. Therefore, the divvy trip for male users from station 283 and station 174 can be for catching up transportation vehicles from their workplaces.

Meanwhile, the top ranked station pair for ``female subscribers'' is ``Station 284 $\to$ Station 255'', where station 284 is next to ``The Art Institute of Chicago'' and station 255 is next to various spots, e.g., ``The Field Museum'', ``Chicago Shedd Aquarium'' and ``Chicago Adler Planetarium''. In addition, between station 284 and station 255, there exist an exercise trail for jogging and bike-riding along the Lake Michigan coast, and Chicago people like to go there for relax a lot. Therefore, the divvy trip for female users from station 284 and station 255 can be for either museum visiting or personal exercises.

For the ``customers'', we observe that the top 5 Divvy trips for them are actually among only 3 stations, which are ``Station 35'' (a station next to Chicago Navy Pier), ``Station 76'' (a station next to Millennium Park) and ``Station 85'' (a station next to the Oak Street Beach). Actually, these 3 stations are all close to attraction spots and are very popular for tourists. In the trip station pair list, we notice that customers will depart and arrive at the same Divvy stations (e.g., Station 76 $\to$ Station 76, and Station 35 $\to$ Station 35), which means they borrow the bike from a station to wander around the nearby places and return the bike back to the same station. However, such observations (i.e., borrowing and returning bikes at common stations) are not common for the subscribed users.




\section{Trip Prediction Problem Formulation}\label{sec:formulation}


Based on the above analysis, we will introduce the trip prediction problem in this section. The trip prediction problem studied in this paper aims at inferring the destination station and trip end time, given that a user borrows a bike from a Divvy station at certain time. We propose to formulate the problem as an origin and destination station pair prediction problem in this paper.


In other words, for a given user, who has borrowed a bike from a known Divvy station $A$ at time $t$, the trip prediction problem aims at returning a set of potential destination Divvy station candidates in the decreasing order of their likelihood that $u$ will ride the bike to as well as the trip duration $\tau$. The trip end time can be represented as $t + \tau$. In the trip prediction problem, we can represent trip origin and destination stations as pairs $(s_o, s_d)$, where $s_o$ denotes the origin station and $s_d$ represents the trip destination station. Based on the existing historical data, a set of features that depict either the user or the characteristics of stations $s_o$, $s_d$ can be extracted, which can be represented as vector $\mb{x}(s_o, s_d) \in \mathbb{R}^k$ of length $k$ (the features will be introduced in Section~\ref{sec:method}). Pair $(s_o, s_d)$ can be labeled with relevance scores between the origin station $s_o$ and the potential destination station $s_d$, which can be represented as $y(s_o, s_d)$ ($y(s_o, s_d) = +1$ if the trip ends at station $s_d$ and $0$ otherwise). Meanwhile, the time duration of trip from $s_o$ and $s_d$ can be denoted as $t(s_o, s_d) \in \mathbb{R}$.

Formally, let $\mathcal{T}$ be the training set containing labeled station pairs. We can represent the features and labels extracted for pairs in $\mathcal{T}$ as $\mathcal{D} = \{\mb{x}(s_o, s_d), y(s_o, s_d), t(s_o, s_d)\}$. The trip prediction problem can be formalized as building two functions $f: \mathbb{R}^k \to \{1, 0\}$ and $h: \mathbb{R}^k \to \mathbb{R}$, where function $f$ maps the station pair feature vector to their inferred relevance score (i.e., the likelihood for the trip to finish at the potential destination stations), while function $h$ maps the feature vector to the inferred trip duration time. These two regression functions will be applied to the potential stations in the test set and can return the predictive confidence scores $\{y(s_o, s_d)\}_{(s_o, s_d)}$ and duration length $\{t(s_o, s_d)\}_{(s_o, s_d)}$ for station pairs in the test set.

\section{Trip Prediction Model}\label{sec:method}

To address the trip prediction problem, in this part, we will introduce the prediction model in detail. First, we will introduce the features extracted for station pairs based on information about users, start time and stations. Next, we will briefly talk about the specific models used in this paper.


\subsection{Features about the user}

Throughout the previous data analysis sections, the type of users (i.e., ``customer'' vs ``male subscriber'' vs ``female subscriber'') have significant influences on the bike trips in both destination stations and trip duration. As a result, based on the user personal information, we propose to extract $3$ features about the users, which include


\begin{itemize}

\item \textit{User Type}: ``Customers'' normally behave very differently from the ``subscribers'' in Divvy bike usage (see Figures in Section~\ref{sec:temporal} and~\ref{sec:spatial}). To differentiate them from each other, based on the user type information, we propose to extract feature $x_1$. If user $u$ is a subscribed user, then $x_1 = +1$; otherwise,  $x_1 = -1$.

\item \textit{User Gender}: ``Male'' uses Divvy bike more often and their activity region concentrates around the Chicago loop area, which is different from the ``female'' users (see Figures~\ref{fig:start}, \ref{fig:end} and \ref{fig:frequent}). To denote the gender about the subscribed users, we define feature $x_2$, where $x_2 = +1$ for ``male subscribers'' and $x_2 = -1$ for ``female subscribers''. For ``customers'', we have no idea about their gender and we will assign $x_2=0$. 


\item \textit{User Age}: In addition, the birth year information is available for subscribers. Young people and mid-aged people tend to use Divvy bike more often (see Figure~\ref{fig:user}). We propose to extract feature $x_3$ to represent the user age. For ``customers'', we set $x_3 = 0$, as we don't know their ages.


\end{itemize}

\subsection{Features about the departure time}

Besides the users information, the Divvy bike usage is also correlated with the trip start time a lot. For instance, people use Divvy bike more often in the summer; ``customers'' tend to use Divvy bike at weekends; ``subscribers'' mainly use the Divvy bike during the rush hours. Therefore, $3$ different features are extracted based on the trip start time:

\begin{itemize}

\item \textit{Month of the trip time}: Winter and early spring in Chicago are not suitable for bike riding (see Figure~\ref{fig:month}). To denote the month of the trip start time, we define feature $x_4$ in the paper, where $x_4 = 1$ if the trip is at January, $x_4 = 2$ if it is at February, and so forth.

\item \textit{Weekday of the trip time}: For subscribed users and customers, they have totally different bike usage patterns on different weekdays (see Figure~\ref{fig:weekday}). To utilize this information, a new feature $x_5$ is introduced. We set $x_5=0$ if the trip starts on Sunday, and set $x_5=1$ for Monday, and so forth.

\item \textit{Hour of the trip time}: Another time-related feature extracted is the specific hour of the start time, as the start hour can show the purpose of the trip a lot (see Figure~\ref{fig:hour}). For simplicity, we divide each day into 24 hours and define another feature $x_6$ to represent the specific trip start hour, where $x_6 = 0$ if it starts within [12AM, 1AM); $x_6=1$ if at [1AM, 2AM); and so forth.

\end{itemize}


These $3$ extracted features show the information about trip start time in $3$ different cyclic patterns.

\subsection{Features about the stations}

In Figure~\ref{fig:frequent}, we have shown the some top frequently commuted station pairs by different categories of users. Therefore, the trip origin station can provide important information to help us infer the destination station as well.  Three different features about the stations are extracted in the experiments:

\begin{itemize}

\item \textit{Station Pairs}: In Figure~\ref{fig:frequent}, we show that some station pairs can be frequently traveled by the users. The first station features extracted for $(s_o, s_d)$ is the station ID pairs, i.e., $x_7 = ID((s_o), ID(s_d))$.


\item \textit{Station geographic information}: Besides the ID information, we also have the geographic information about the stations, which can be represented as the (latitude, longitude) pairs. The coordinate pairs are also used as a feature, which can be represented as $x_8 = \mbox{latitude}(s_o), \mbox{longitude}(s_o), \mbox{latitude}(s_d), \mbox{longitude}(s_d)$.

\item \textit{Geographic Distance}: The majority of Divvy bike trips are of length $0.5-5$KM (see Figure~\ref{fig:trip_distance}) and trips that are too short (around $0$KM) or too long (longer than $10$KM) are very rare. We propose to extract feature $x_9$ to denote geographic distance between stations $(s_o, s_d)$, where \textit{Manhattan Distance} is used as the distance measure.

\end{itemize}

Based on the above extracted features, we can represent the feature vector for certain station pairs $(s_o, s_d)$ as $\mb{x}(s_o, s_d) = [x_1, x_2, \cdots, x_9]$ of length $13$ in total (as $x_7$ and $x_8$ are of lengths 2 and 4 respectively), which together with the label $y(s_o, s_d)$ and time $t(s_o, s_d)$ can be used to build the confidence score and trip duration prediction models.

\subsection{Trip Destination Station Inference Model}

For the trip destination station prediction problem, we propose to map it to a binary classification in the experiments. For each trip origin and destination station pairs (e.g., $(s_o, s_d)$), we assign it with different labels $\{1, 0\}$ to denote whether a certain trip starting at $s_o$ will end at $s_d$ or not. To address the problem, we propose to apply a state-of-the-art pairwise based regression algorithm, namely MART (Multiple Additive Regression Trees) \cite{WBSG10}, to develop a regression function. MART is based on the stochastic gradient boosting approach described in \cite{F00, F02} which performs gradient descent optimization in the functional space. In our experiments, we used the log-likelihood as the loss function, steepest-descent (gradient descent) as the optimization technique, and binary decision trees as the fitting function. For more information about the MART model, please refer to \cite{F00, F02, TWAP11}.

\subsection{Trip Duration Inference Model}

To predict the time length of the trip, the same set of features are applied to build the trip duration inference model. Different regression models can be used as the base prediction model, and, without a loss of generality, we will apply the Lasso regression model as the base regression model in this paper, which fits a linear equation $\hat{t}(s_o, s_d) = \sum_{i = 1}^k b_i x_i + b_0$, where $\hat{t}(s_o, s_d)$ is the inferred trip length between stations $s_o$ and $s_d$, term $b_i$ denotes the coefficient of feature $x_i$ and $b_0$ represents the bias term.

To get the coefficient values in training the model, Lasso uses the $L_1$ prior as the regularizer, and the optimal coefficients can be learned by solving the following equation
$$\arg \min_{\mb{b}} \sum_{(s_o, s_d)} (\hat{t}(s_o, s_d) - t(s_o, s_d))^2 + \alpha \left | \mb{b} \right |_1,$$
where $t(s_o, s_d)$ is the real duration of the trip between $s_o$ and $s_d$ and $\alpha$ is the weight of the regularizer term.

\section{Experiments}\label{sec:exp}

To test the effectiveness of these two models in addressing the trip prediction problem, we conduct experiments on the real-world bicycle-sharing system Divvy (introduced in Section~\ref{sec:data}). In this section, we will first introduce the experiment settings, which include the experiment setups, comparison methods and evaluation metrics. Next, we will show the experiment results and give detailed analysis.

\begin{figure*}[t]
\centering
\subfigure[Accuracy]{ \label{fig:accuracy}
    \begin{minipage}[l]{.45\columnwidth}
      \centering
      \includegraphics[width=\textwidth]{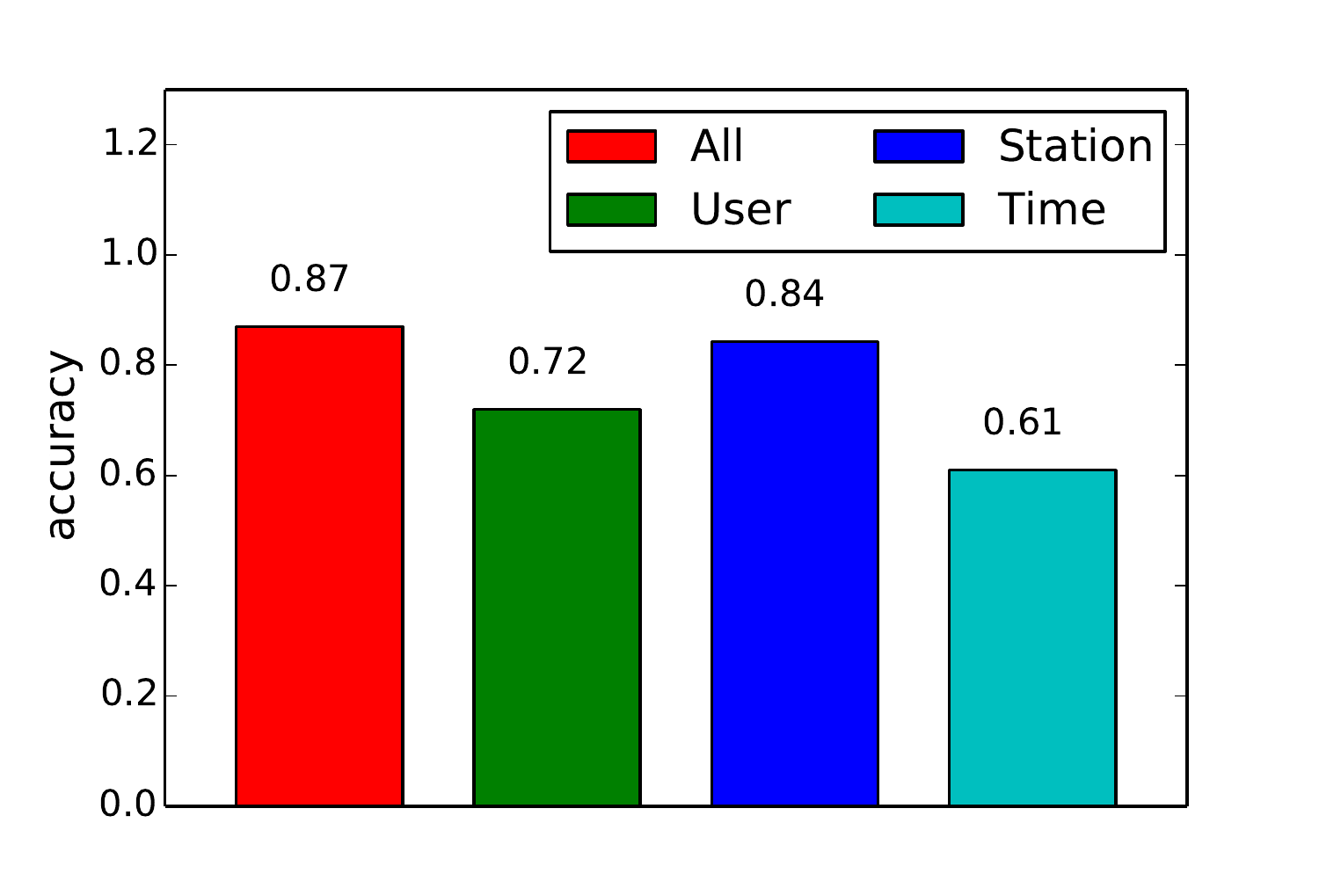}
    \end{minipage}
  }
  \subfigure[Precision]{\label{fig:precision}
    \begin{minipage}[l]{.45\columnwidth}
      \centering
      \includegraphics[width=\textwidth]{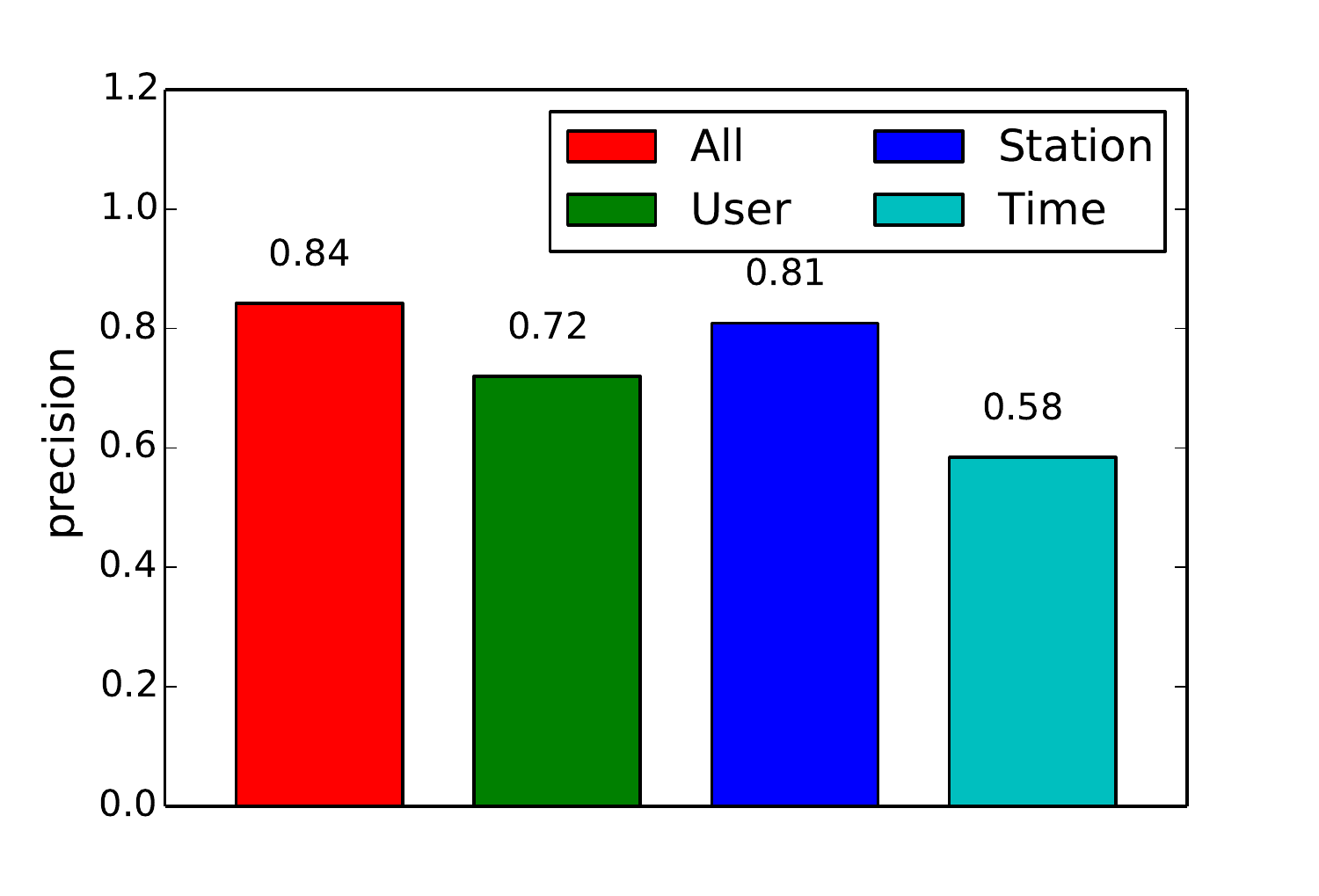}
    \end{minipage}
  }
  \subfigure[Recall]{\label{fig:recall}
    \begin{minipage}[l]{.45\columnwidth}
      \centering
      \includegraphics[width=\textwidth]{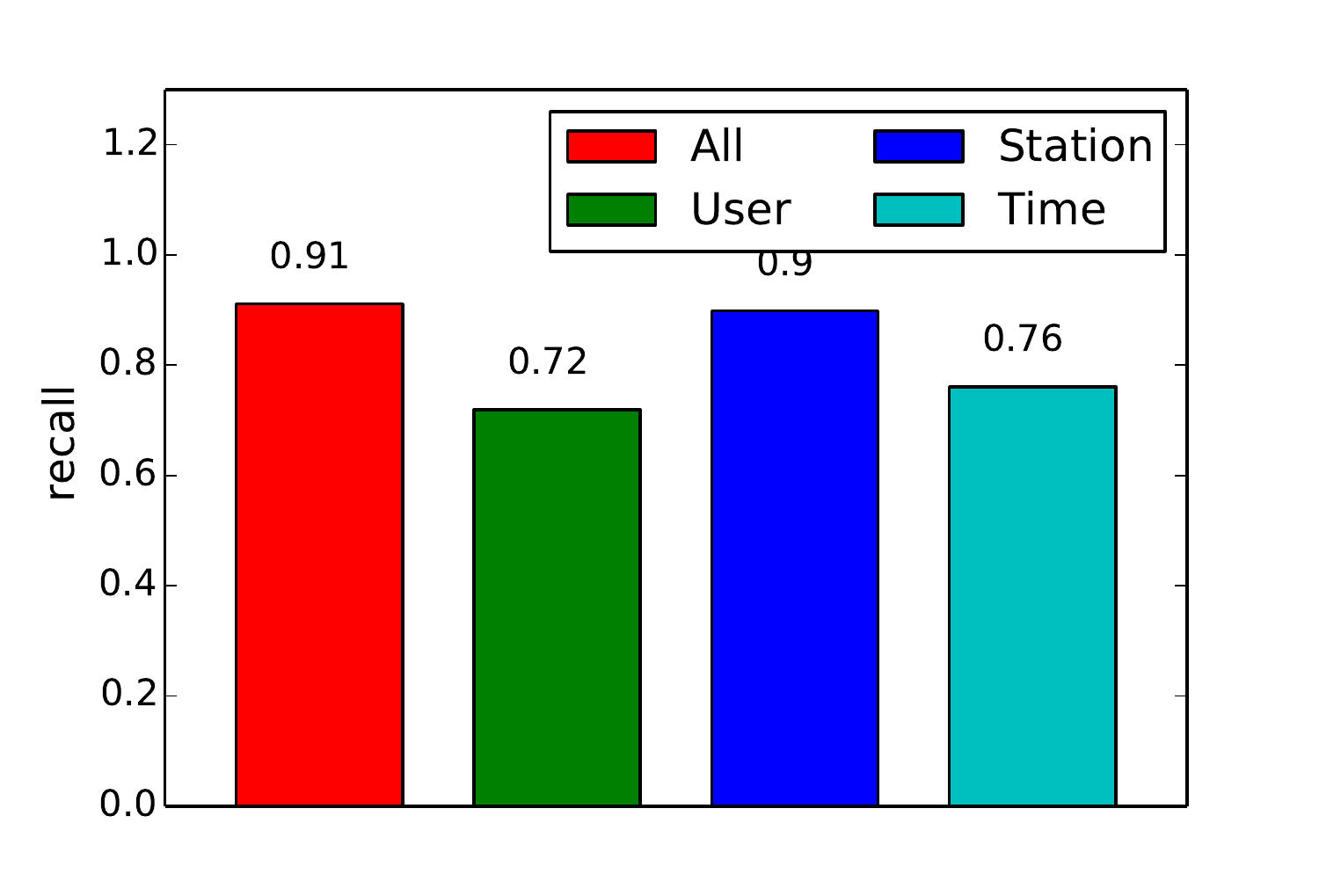}
    \end{minipage}
  }
  \subfigure[F1]{\label{fig:f1}
    \begin{minipage}[l]{.45\columnwidth}
      \centering
      \includegraphics[width=\textwidth]{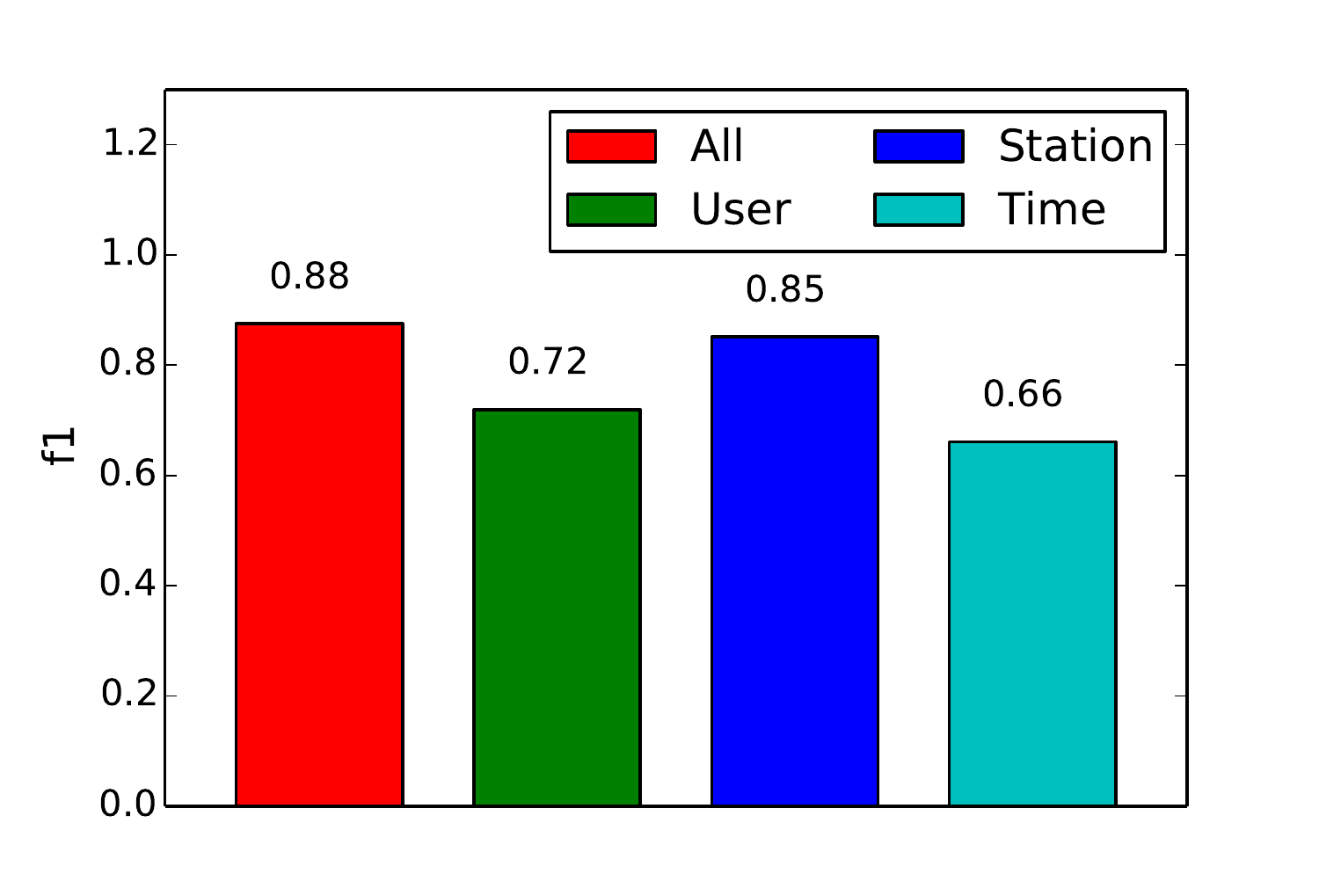}
    \end{minipage}
  }
\vspace{-5pt}
  \caption{Trip destination station prediction results evaluated by different metrics.}\label{fig:end_station_result}\vspace{-15pt}
\end{figure*}

\subsection{Experiment Settings}

\subsubsection{Experiment Setups}
From the dataset, we extract the trip tuples $(\mbox{user}, \mbox{origin station}, \mbox{destination station}, \mbox{departure time})$ as the existing trip set, where each tuple contains the complete information about the trip. In the trip destination station prediction problem, the existing trip set is used as the positive set (i.e., assigned with labels $1$), and a equal-sized negative trip tuple set is random generated, instances in which are assigned with labels $0$. In the negative tuple set, (1) users can be ``customers'' and ``subscribers'' of equal chance, and the gender of ``subscribers'' is randomly assigned with either ``male'' and ``female'', whose ages are random selected from $\{1, 2, \cdots, 100\}$; (2) the origin and destination are randomly selected from the whole station set; and (3) the negative trip departure time is random selected from July $1st$, 2013 to June $30th$, 2015. Both positive and negative trip sets are divided into two parts according to ratio $4:1$ based on the time order, where $4$ folds are used as the training set and $1$ fold is used as the test set. A set of features are extracted for each instance in the training and test sets. We train the trip end prediction model MART with the training set, which will be applied to the test set to infer the labels of the test pairs. 


Meanwhile, in the trip duration inference problem, similarly, we divide the existing trip set into two parts according to ratio $4:1$ based on the time order, where $4$ folds are used as the training set and $1$ fold is used as the test set. However, the setting of trip duration inference is slightly different: (1) no negative trip set is needed; and (2) the instances in the training and test set are assigned with their real-trip duration as their labels. The same set of features are extracted to build the trip duration inference model (i.e., Lasso) based on the training set, which will be applied to infer the trip time duration of instances in the test set. 

\subsubsection{Comparison Methods}

The comparison methods used in trip destination and duration inference can be divided into two categories depending on the information used:

\noindent \textbf{Models using all information}

\begin{itemize}

\item \textit{{\our}}: Method {\our} builds the trip end prediction and trip duration inference models with all the three categories of features extracted, which include \textit{user}, \textit{station} and \textit{time} based features.

\end{itemize}

\noindent \textbf{Models using partial information}

\begin{itemize}

\item \textit{{\ouruser}}: Method {\ouruser} builds the trip end prediction and trip duration inference models with the features about \textit{users} only.

\item \textit{{\ourstation}}: Method {\ourstation} builds the trip end prediction and trip duration inference models with the features about \textit{stations} only.

\item \textit{{\ourtime}}: Method {\ourtime} only uses the features about the \textit{time} only to build the trip end prediction and trip duration inference models.

\end{itemize}

\subsubsection{Evaluation Metrics}

To evaluate the performance of these different methods in addressing the trip prediction problem, we apply different evaluation metrics to measure their prediction results.

We formulate the trip end prediction problem as a binary classification problem, and all these $4$ comparison methods can output the predicted labels of trip pairs in the test set. By comparing them with the ground-truth labels, we can evaluate their performance with $4$ frequently used metrics: Accuracy, Precision, Recall and F1-score. 

We formulate the trip duration inference problem as a regression problem, and the comparison methods will output the inferred the time duration of trips in the test set. Meanwhile, we also have the real-world trip duration from the dataset, i.e., the ground-truth. Different metrics used for regression problems can be applied here, and we use the MAE (Mean Absolute Error) and $R^2$ (i.e., Coefficient of Determination) as the evaluation metrics.

\begin{figure}[t]
\centering
\subfigure[MAE]{ \label{fig:mae}
    \begin{minipage}[l]{.45\columnwidth}
      \centering
      \includegraphics[width=\textwidth]{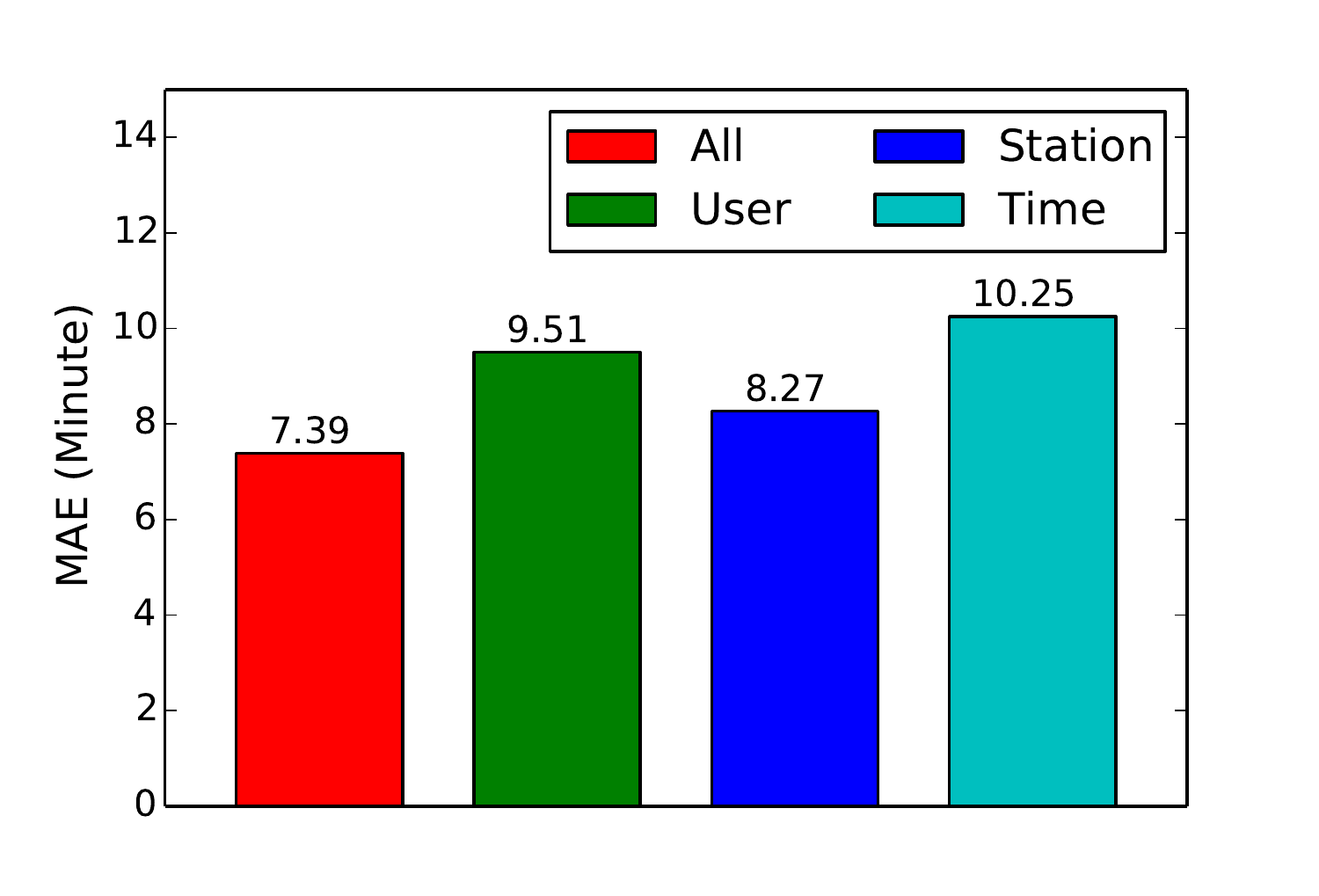}
    \end{minipage}
  }
  \subfigure[$R^2$]{\label{fig:r2}
    \begin{minipage}[l]{.45\columnwidth}
      \centering
      \includegraphics[width=\textwidth]{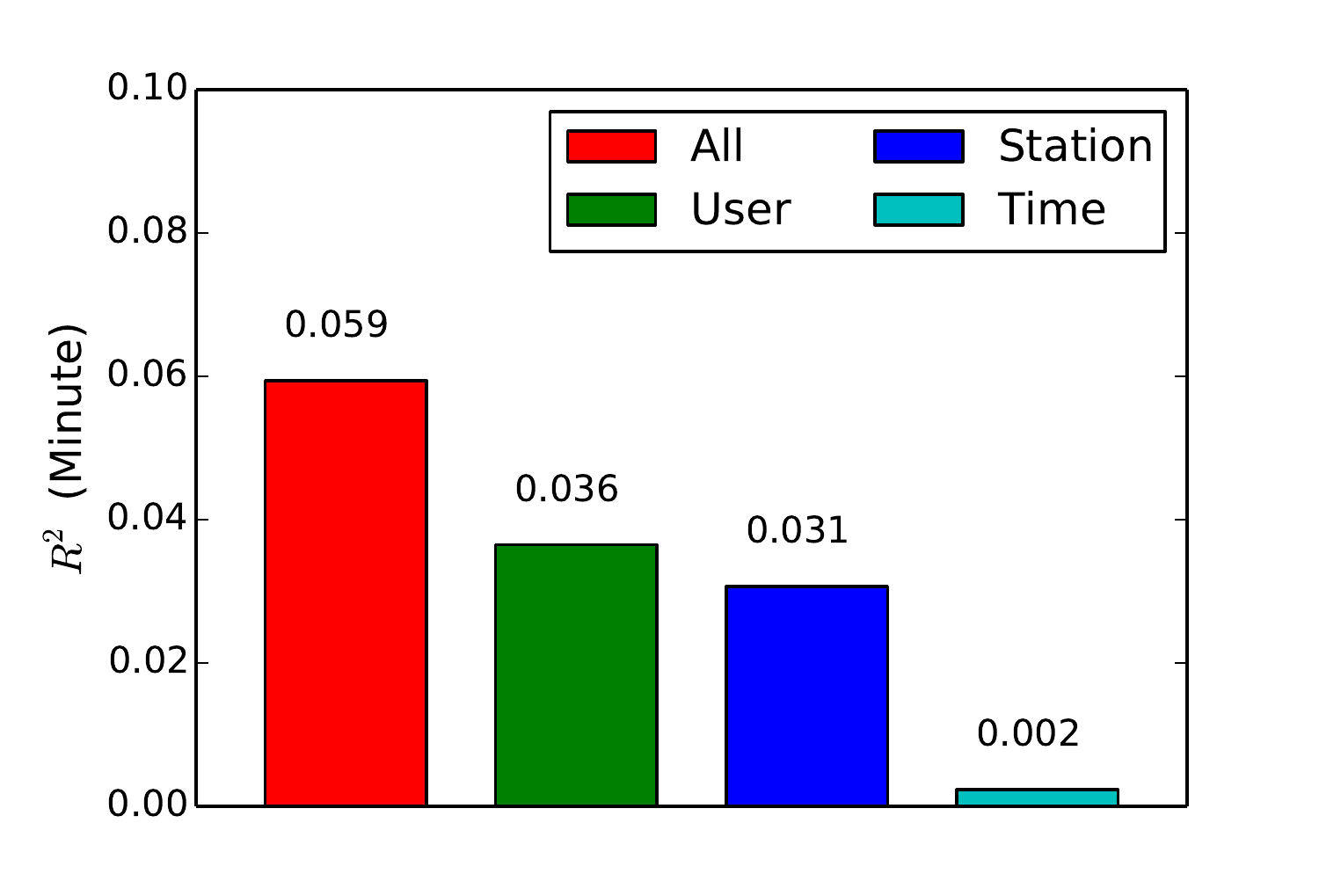}
    \end{minipage}
  }
\vspace{-5pt}
\caption{Trip duration inference results.}\label{fig:duration_result}\vspace{-15pt}
\end{figure}

\subsection{Experiment Results}

The experiment results are available in Figure~\ref{fig:end_station_result} and Figure~\ref{fig:duration_result}. Figure~\ref{fig:end_station_result} show the results of trip end prediction and Figure~\ref{fig:duration_result} gives the results of trip duration inference.


By comparing {\our} with the other methods in Figure~\ref{fig:end_station_result}, we can observe that {\our} can outperform other methods with significant advantages consistently evaluated by different metrics. For instance, in Figure~\ref{fig:accuracy}, the Accuracy achieved by {\our} is $0.87$ which is about $21\%$ higher than the Accuracy gained by {\ouruser} (i.e., $0.72$); $4\%$ higher than the Accuracy score achieved by {\ourstation} (i.e., $0.84$); and $43\%$ larger than the Accuracy score obtained by {\ourtime} (i.e., $0.61$). Similar results can be observed in Figures~\ref{fig:precision}-\ref{fig:f1}, where Precision, Recall and F1 are used as the evaluation metrics.

Generally, among the comparison methods, {\our} utilizing all these $3$ categories of features perform the best. Among the $3$ methods using one category of feature only, {\ourstation} can outperform {\ouruser}, while {\ouruser} performs better than {\ourtime}. It is also easy to understand, as the task is to infer the trip destination station, historical trip station pair information can only provide more direct information for addressing the task. Meanwhile, the features about users and trip start time can provide the indirect hints, as they are about the bike user and time, not directly about the stations.

In Figure~\ref{fig:mae}-\ref{fig:r2}, we show the results about trip duration inference problem, which are evaluated by both MAE and $R^2$ metrics. Compared with other comparison methods, {\our} achieves better performance with the much smaller MAE and larger $R^2$ score. For example, the MAE introduced by {\our} is $7.39$ (minute), which is $22.23\%$ lower that the MAE introduced by {\ouruser}, $10.64\%$ smaller than the MAE introduced by {\ourstation} and $28\%$ lower than the MAE introduced by {\ourtime}. For the $R^2$ metric, the $R^2$ score achieved by {\our} is $0.059$, which is nearly the double of the $R^2$ scores gained by {\ouruser} and {\ourstation}. The advantages of {\our} against {\ourtime} is more obvious: the MAE of {\our} accounts for only $72\%$ of that achieved by {\ourtime}; and the $R^2$ score obtained by {\our} is as large as the 30 times of the $R^2$ achieved by {\ourtime}.

Therefore, by utilizing the complete information available about the trips, {\our} can outperform other comparison methods with significant advantages in both predicting the trip destination stations and inferring the trip duration time.

\section{Related Work} \label{sec:related}

Bicycle-sharing has received increasing attention in recent years with initiatives to increase cycle usage improve the  first mile/last mile connection to other modes of transit, and lessen the environmental impacts of our transport activities. DeMaio gives a complete introduction about the history, impacts, models of provision, and future of bicycle-sharing systems in \cite{D09}. Midgley provides a complete overview work about the bicycle-sharing schemes, management, policies, and challenges as well as opportunities in \cite{M11}. A large number of other review and case-study works on bicycle-sharing systems have appeared so far \cite{M09, ZZDB15, VM15, EL14}, which study the bicycle-sharing systems from different aspects and directions.

Recently, urban computing has become a hot research area and lots of works have been done by Zheng et al. already \cite{ZYLLSCL15, PZWS13, LZZC15}. The bicycle-sharing systems are an important part in urban computing. Many research works have been done on bicycle-sharing systems and other transportation systems to study the system design problem \cite{LYC13}, load balance problem \cite{PSFR13}, and bicycle traffic prediction problem \cite{LZZC15}. Lin et al. \cite{LYC13} introduce a strategic design problem for bicycle sharing systems incorporating bicycle stock considerations, which is formulated as a hub location inventory model. The problem studied in \cite{LYC13} covers the design work about various aspects of the bicycle-sharing system, e.g., the number and locations of bicycle stations, the creation of bicycle lanes, the selection of paths, etc. Pavone et al. develop methods for maximizing the throughput of a mobility-on-demand urban transportation system and introduce a rebalancing policy that minimizes the number of vehicles performing rebalancing trips \cite{PSFR13}. The optimal rebalancing policy can be found as the solution to a linear program effectively in the proposed model. Li et al. propose a hierarchical prediction model to predict the number of bikes that will be rent from/returned in a future period for bicycle-sharing systems \cite{LZZC15}, which focus more on the macroscopic bike traffic flow in the bicycle-sharing system and is different from the microscopic trip destination and duration prediction problem of a specific trip studied in this paper.

\section{Conclusion}\label{sec:conclusion}

In this paper, we have studied the trip prediction problem for bicycle-sharing systems to infer the potential trip destination station and trip duration. Extensive analysis about the user composition of a real-world bicycle-sharing system, individuals' temporal bike usage behavior patterns and spatial bike usage behavior patterns have been done. Based on the analysis results, two new regression based inference models have been introduced in this paper to predict the potential trip destination station and trip duration respectively. Experiments conducted on the real-world bicycle-sharing system dataset demonstrate the effectiveness of the proposed model.
\label{sec:ack}
\section{Acknowledgement}

This work is supported in part by NSF through grants III-1526499, and CNS-1115234,  and Google Research Award.

This research is also partially supported by the grant from the Natural Science Foundation of China (No. 61303017), the Natural Science Foundation of Hebei Province (No. F2014210068).
\balance
\bibliographystyle{plain}
\bibliography{reference}

\begin{thebibliography}{10}

\bibitem{chicago}
Divvy: Chicago's newest transit system.
\newblock
  \url{http://divvybikes.tumblr.com/post/95362930905/expansion-map-2015}.
\newblock [Online; accessed 24-November-2015].

\bibitem{CSN09}
A.~Clauset, C.~Shalizi, and M.~Newman.
\newblock Power-law distributions in empirical data.
\newblock {\em SIAM Review}, 2009.

\bibitem{D09}
P.~DeMaio.
\newblock Bike-sharing: History, impacts, models of provision, and future.
\newblock {\em Journal of Public Transportation}, 2009.

\bibitem{EL14}
C.~Etienne and O.~Latifa.
\newblock Model-based count series clustering for bike sharing system usage
  mining: A case study with the v\'{E}lib system of paris.
\newblock {\em ACM Transaction Intelligent Systems and Technology}, 2014.

\bibitem{F00}
J.~Friedman.
\newblock Greedy function approximation: A gradient boosting machine.
\newblock {\em Annals of Statistics}, 2000.

\bibitem{F02}
J.~Friedman.
\newblock Stochastic gradient boosting.
\newblock {\em Computational Statistics \& Data Analysis}, 2002.

\bibitem{LZZC15}
Y.~Li, Y.~Zheng, H.~Zhang, and L.~Chen.
\newblock Traffic prediction in a bike-sharing system.
\newblock In {\em SIGSPATIAL}, 2015.

\bibitem{LYC13}
J.~Lin, T.~Yang, and Y.~Chang.
\newblock A hub location inventory model for bicycle sharing system design:
  Formulation and solution.
\newblock {\em Computers and Industrial Engineering}, 2013.

\bibitem{M09}
P.~Midgley.
\newblock The role of smart bike-sharing systems in urban mobility.
\newblock {\em Land Transport Authority JOURNEY}, 2009.

\bibitem{M11}
P.~Midgley.
\newblock Bicycle-sharing schemes: Enhancing sustainable mobility in urban
  areas.
\newblock {\em United Nations Department of Economic and Social Affairs}, 2011.

\bibitem{PZWS13}
B.~Pan, Y.~Zheng, D.~Wilkie, and C.~Shahabi.
\newblock Crowd sensing of traffic anomalies based on human mobility and social
  media.
\newblock In {\em SIGSPATIAL}, 2013.

\bibitem{PSFR13}
M.~Pavone, S.~Smith, E.~Frazzoli, and D.~Rus.
\newblock Load balancing for mobility-on-demand systems.
\newblock {\em International Journal of Robotics Research}, 2012.

\bibitem{TWAP11}
S.~Tyree, K.~Weinberger, K.~Agrawal, and J.~Paykin.
\newblock Parallel boosted regression trees for web search ranking.
\newblock In {\em WWW}, 2011.

\bibitem{VM15}
P.~Vogel and D.~Mattfeld.
\newblock Strategic and operational planning of bike-sharing systems by data
  mining – a case study.
\newblock {\em Computational Logistics}, 2015.

\bibitem{WM97}
D.~Wilson and T.~Martinez.
\newblock Improved heterogeneous distance functions.
\newblock {\em Journal of Artificial Intelligence Research}, 1997.

\bibitem{WBSG10}
Q.~Wu, C.~Burges, K.~Svore, and J.~Gao.
\newblock Adapting boosting for information retrieval measures.
\newblock {\em Information Retrieval}, 2010.

\bibitem{ZZDB15}
L.~Zhang, J.~Zhang, Z.~Duan, and D.~Bryde.
\newblock Sustainable bike-sharing systems: Characteristics and commonalities
  across cases in urban china.
\newblock {\em Journal of Cleaner Production}, 2015.

\bibitem{ZYLLSCL15}
Y.~Zheng, X.~Yi, M.~Li, R.~Li, Z.~Shan, E.~Chang, and T.~Li.
\newblock Forecasting fine-grained air quality based on big data.
\newblock In {\em KDD}, 2015.

\end{thebibliography}

\end{document}